\documentclass[11pt,a4paper]{article}

\usepackage[utf8]{inputenc}
\usepackage{jheppub}

\usepackage{lmodern}
\usepackage[T1]{fontenc}

\usepackage{amsmath}
\usepackage{graphicx}
\usepackage{url}
\usepackage{epsfig}
\usepackage{float}
\usepackage{epsfig}
\usepackage{amssymb}
\usepackage{amsmath}
\usepackage{url}
\usepackage{hyperref}
\usepackage{color}
\usepackage{cancel}
\usepackage{array}
\usepackage{multirow}
\usepackage{mathrsfs}

\newcommand{\squ}{{\tilde{q}}}

\newcommand{\glu}{{\tilde{g}}}
\newcommand{\sto}{{\tilde{t}}}

\newcommand{\totd}{\mathrm{d}}
\newcommand{\mav}{m_{\mathrm{av}}}
\newcommand{\mred}{m_{\mathrm{red}}}

\newcommand{\refeq}[1]{eq.~\eqref{#1}}
\newcommand{\refeqs}[1]{eqs.~\eqref{#1}}
\newcommand{\Refeq}[1]{Eq.~\eqref{#1}}

\newcommand{\reffig}[1]{figure~\ref{#1}}

\newcommand{\Reffig}[1]{Figure~\ref{#1}}

\newcommand{\alphas}{\alpha_{\mathrm{s}}}

\title{NNLL-fast: predictions for coloured supersymmetric particle production at the LHC with threshold and Coulomb resummation}

\author[a]{Wim Beenakker,}
\author[b]{Christoph Borschensky,}
\author[c]{Michael Kr\"amer,}
\author[d]{Anna Kulesza,}
\author[e]{Eric Laenen}

\affiliation[a]{Theoretical High Energy Physics, IMAPP, Faculty of Science, Mailbox 79, Radboud University Nijmegen, P.O. Box 9010, NL-6500 GL Nijmegen,\\
Institute of Physics, University of Amsterdam, Amsterdam, The Netherlands}
\affiliation[b]{Institute for Theoretical Physics, University of T\"ubingen, Auf der Morgenstelle 14, 72076 T\"ubingen, Germany}
\affiliation[c]{Institute for Theoretical Particle Physics and Cosmology, RWTH Aachen University D-52056 Aachen, Germany}
\affiliation[d]{Institute for Theoretical Physics, WWU M\"unster, D-48149 M\"unster, Germany}
\affiliation[e]{ITFA, University of Amsterdam, Science Park 904, 1018 XE, Amsterdam,\\
ITF, Utrecht University, Leuvenlaan 4, 3584 CE Utrecht,\\
Nikhef Theory Group, Science Park 105, 1098 XG Amsterdam, The Netherlands}

\emailAdd{w.beenakker@science.ru.nl}
\emailAdd{christoph.borschensky@uni-tuebingen.de}
\emailAdd{mkraemer@physik.rwth-aachen.de}
\emailAdd{anna.kulesza@uni-muenster.de}
\emailAdd{t45@nikhef.nl}

\abstract{
We present state-of-the art predictions for the production of supersymmetric squarks and gluinos at the 
Large Hadron Collider (LHC), including soft-gluon resummation up to next-to-next-to-leading logarithmic (NNLL) accuracy, 
the resummation of Coulomb corrections and the contribution from bound states. The NNLL corrections enhance the 
cross-section predictions and reduce the scale uncertainty to a level of 5-10\%. The NNLL resummed cross-section  
predictions can be obtained from the computer code \texttt{NNLL-fast}, which also provides the scale uncertainty and 
the pdf and $\alpha_{\rm s}$ error.}


\begin{document}
\begin{flushright}
MS-TP-16-17\\Nikhef/2016-037\\TTK-16-28
\end{flushright}
\maketitle

\renewcommand{\arraystretch}{1.2}

\section{Introduction}\label{s:intro}
One of the most important goals of the present phase of the Large Hadron 
Collider (LHC) programme is the search for supersymmetry (SUSY) 
\cite{Wess:1973kz,Wess:1974tw,Fayet:1976et,Farrar:1978xj,Sohnius:1985qm,Martin:1997ns}. 
SUSY predicts a rich variety of new particles and a number of 
dark-matter candidates, while at the same time providing a natural solution to 
the hierarchy problem and facilitating gauge-coupling unification at a high 
energy scale. For SUSY to be natural, it has to be realised close to the 
electroweak scale, within the reach of the current Run II of the LHC with a 
centre-of-mass energy of $\sqrt{S} = 13$ TeV. In particular, the partners of 
the coloured particles -- squarks ($\squ$) and gluinos ($\glu$) -- would be 
produced in copious amounts if their masses were around a few TeV or below. Experimental
searches set a current lower mass limit on the coloured supersymmetric 
particles of around 1 TeV up to 1.8 TeV, depending on the specific SUSY model
\cite{Aad:2015iea, Aad:2015baa, Aad:2015pfx, Chatrchyan:2014goa, Khachatryan:2015lwa,Khachatryan:2016pup}.

In the framework of the Minimal Supersymmetric Standard Model (MSSM) with 
\mbox{R-parity} conservation~\cite{Nilles:1983ge,Haber:1984rc}, supersymmetric 
particles (or short: sparticles) are always produced in pairs at collider 
experiments. In the following we therefore consider pair-production of squarks 
and gluinos through the collision of two hadrons $h_1$ and $h_2$. i.e.
\begin{align*}
h_1h_2 \to \squ\squ^*, \; \squ\squ\;(+\,\squ^*\squ^*)\,, \; \glu\glu\,, 
           \; \squ\glu\;(+\,\squ^*\glu) +X
\qquad\text{and}\qquad
h_1h_2 \to \sto_1\sto_1^*\,, \; \sto_2\sto_2^* +X\,.
\end{align*}
In the first set of production channels, charge-conjugated processes are 
combined and the chiralities of the squarks $(\squ_L,\squ_R)$ are understood to
be summed over. Also the different squark flavours  are understood to be summed
over with the exception of the stops ($\sto$), the superpartners of the 
third-generation top quarks. The production of stops is treated separately because of the potentially strong mixing between the weakly-interacting gauge 
eigenstates of the stops, which can lead to a large mass splitting between the 
lighter ($\sto_1$) and heavier ($\sto_2$) physical stop mass 
eigenstates~\cite{Ellis:1983ed} and to a stronger dependence on SUSY model
parameters. 

Accurate predictions for the production of squarks and gluinos are important to
derive mass exclusion limits, and, should SUSY be realised in nature, to 
precisely measure the masses and properties of the sparticles. 
The next-to-leading order (NLO) corrections to total production cross 
sections and differential distributions have been known for some time, both within QCD 
\cite{Beenakker:1994an,Beenakker:1995fp,Beenakker:1996ch,Beenakker:1997ut, GoncalvesNetto:2012yt, Hollik:2013xwa, Gavin:2013kga, Hollik:2012rc, Gavin:2014yga, Degrande:2015vaa}
and including electroweak corrections \cite{Hollik:2007wf,Beccaria:2008mi,Hollik:2008yi,Hollik:2008vm,Mirabella:2009ap,Hollik:2010vn, Germer:2014jpa,Hollik:2015lha}. 
An important contribution to the total cross section originates from the 
kinematical regime where the production of the coloured sparticles proceeds 
close to the production threshold. 
In particular for high sparticle masses, the production cross sections receive 
a significant fraction of the NLO corrections from this kinematical regime due 
to soft-gluon emission off the coloured initial- and final-state particles. 
Furthermore, Coulombic gluon exchange between the slowly moving heavy 
final-state particles results in additional large contributions. 

Taking into account the soft-gluon corrections to all orders in the strong 
coupling can be achieved by means of threshold resummation techniques in
Mellin space \cite{Sterman:1986aj,Catani:1989ne,Bonciani:1998vc,Contopanagos:1996nh,Kidonakis:1998bk,Kidonakis:1998nf}. In this framework, corrections from 
soft gluons at next-to-leading logarithmic (NLL) accuracy have already been 
available for a while for all production processes of squarks and gluinos in 
the MSSM, including stops \cite{Kulesza:2008jb,Kulesza:2009kq,Beenakker:2009ha,Beenakker:2010nq,Beenakker:2011fu}. 
Alternatively, the analysis of the soft-gluon effects can be  
performed in the framework of soft-collinear effective theory (SCET), which also allows an efficient treatment of
Coulomb contributions \cite{Beneke:2010da,Falgari:2012hx}. 
The present state-of-the-art is that threshold resummation corrections up to 
next-to-next-to-leading logarithmic (NNLL) accuracy have been implemented for 
gluinos and first- and second-generation squarks in 
\cite{Beenakker:2011sf,Langenfeld:2012ti,Pfoh:2013edr,Beenakker:2013mva,Beneke:2013opa,Beneke:2014wda,Beenakker:2014sma}, 
and recently also for stops in \cite{Broggio:2013cia,Beenakker:2016gmf}.

Coulombic gluon-exchange effects have first been studied for the case of 
top quark production at $e^+e^-$ colliders 
\cite{Fadin:1990wx,Kwong:1990iy,Strassler:1990nw}.
As they solely depend on the colour structure of the final-state particles, the
methods derived for top production can straightforwardly be applied to other 
processes like the production of squarks and gluinos. While resummed Coulomb 
contributions involving a leading-order Coulomb interaction can be described by
a Sommerfeld factor (see e.g.\ \cite{Fadin:1990wx}), higher-order corrections 
to the Coulomb potential can be conveniently implemented through Green's 
functions in the non-relativistic QCD (NRQCD) approach \cite{Fadin:1987wz,Hoang:2000yr,Beneke:1999qg}. 

In this work, we improve the NNLL predictions for all squark and gluino 
production cross sections obtained with the conventional Mellin-space 
technique by including resummation of Coulomb corrections and by adding effects
from bound states. We analyse the impact of these types of corrections on the 
size of the cross section and its theoretical error. The effects of NNLL 
resummation on squark and gluino production processes are demonstrated at
the LHC with $\sqrt S=13$ TeV, examining squark and gluino masses up
to 3~TeV. Our results can be obtained from a new computer code, 
called \texttt{NNLL-fast}~\cite{nnll-fast}, which provides NNLL predictions and error estimates for these processes.

The structure of the paper is as follows. In section~\ref{s:theoryresum} we briefly review the NNLL resummation formalism in Mellin space, 
including in particular our treatment of the Coulomb and bound-state corrections. Numerical results are presented in section~\ref{s:numerics}, 
where we also compare with results obtained within the SCET formalism. We conclude in section~\ref{s:conclusion}. 

\section{NNLL formalism in Mellin space: a brief review}
\label{s:theoryresum}

We begin with the general formulation of the resummed total cross section in 
Mellin space for hadronic processes involving the production of a heavy 
particle pair. Subsequently we move on to discuss the resummation of 
Coulomb corrections and the inclusion of bound-state effects.

\subsection{Resummation of threshold logarithms at NNLL}

 The inclusive hadronic cross section for the production of particles $k$ and 
$l$, $\sigma_{h_1h_2\to kl}$, can be written in terms of the partonic cross 
section, $\sigma_{ij\to kl}$, in the following manner:
\begin{multline}
  \label{eq:hadr-cross}
  \sigma_{h_1 h_2 \to kl}\bigl(\rho, \{m^2\}\bigr) 
  \;=\; \sum_{i,j} \int d x_1 d x_2\,d\hat{\rho}\;
        \delta\left(\hat{\rho} - \frac{\rho}{x_1 x_2}\right)\\
        \times\,f_{i/h_{1}}(x_1,\mu^2 )\,f_{j/h_{2}}(x_2,\mu^2 )\,
        \sigma_{ij \to kl}\bigl(\hat{\rho},\{ m^2\},\mu^2\bigr)\,.
\end{multline}
Here $\{m^2\}$ denotes all the masses entering the calculation, $i$ and $j$ are 
the initial-state parton flavours, $f_{i/h_{1}}$ and $f_{j/h_{2}}$ are the parton 
distribution functions, $\mu$ is the common factorisation and renormalisation 
scale, $x_1$ and $x_2$ are the momentum fractions of the partons inside the 
hadrons $h_1$ and $h_2$, and $\rho$ and $\hat{\rho}$ are the hadronic and
partonic threshold variables, respectively. The threshold for the production of 
two final-state particles $k$ and $l$ with masses $m_k$ and $m_l$ corresponds to
a hadronic center-of-mass energy squared of $S={(m_k+m_l)}^2$. Therefore we 
define the hadronic threshold variable $\rho$, measuring the distance from 
threshold in terms of a quadratic energy fraction, as
\[\rho \;=\; \frac{{(m_k+m_l)}^2}{S} \;=\; \frac{4(m_{\rm av})^2}{S}\,,\]
where $m_{\rm av}=(m_k+m_l)/2$ is the average mass of the final-state particles $k$ 
and $l$.

In the threshold regime, the dominant contributions to the higher-order QCD 
corrections due to soft-gluon emission have the general form
\begin{equation}
\alpha_{\rm s}^n \log^m\!\beta^2\ \ , \ \ m\leq 2n 
\qquad {\rm \ with\ } \qquad 
\beta^2 \,\equiv\, 1-\hat{\rho} \,=\, 1 \,-\, \frac{4(m_{\rm av})^2}{s}\,,
\label{eq:beta}
\end{equation}
where $s=x_1x_2S$ is the partonic center-of-mass energy squared and 
$\alpha_{\rm s}$ the strong coupling. We perform the resummation of the 
soft-gluon emission after taking the Mellin transform (indicated by a tilde) of
the cross section:
\begin{align}
  \label{eq:Mellin-transf}
  \tilde\sigma_{h_1 h_2 \to kl}\bigl(N, \{m^2\}\bigr) 
  &\equiv \int_0^1 d\rho\;\rho^{N-1}\;
           \sigma_{h_1 h_2\to kl}\bigl(\rho,\{ m^2\}\bigr) \nonumber\\[2mm]
  &=      \;\sum_{i,j} \,\tilde f_{i/{h_1}} (N+1,\mu^2)\,
           \tilde f_{j/{h_2}} (N+1, \mu^2) \,
           \tilde{\sigma}_{ij \to kl}\bigl(N,\{m^2\},\mu^2\bigr)\,.
\end{align}
The logarithmically enhanced terms now take the form of 
$\alpha_{\rm s}^n \log^m\!N$, $m\leq 2n$, where the threshold limit 
$\beta \rightarrow 0$ corresponds to $N \to \infty$. The all-order summation of 
such logarithmic terms follows from the near-threshold factorisation of the 
cross section into functions that each capture the contributions of classes of 
radiation effects: hard, collinear and wide-angle soft radiation 
\cite{Sterman:1986aj,Catani:1989ne,Bonciani:1998vc,Contopanagos:1996nh,Kidonakis:1998bk,Kidonakis:1998nf}. 
Near threshold the resummed partonic cross section takes the form
\begin{align}
  \label{eq:resummed-cross}
  \tilde{\sigma}^{\rm (res)} _{ij\to kl}\bigl(N,\{m^2\},&\mu^2\bigr) 
  = \sum_{I}\,\tilde\sigma^{(0)}_{ij\to kl,I}\bigl(N,\{m^2\},\mu^2\bigr)\, 
    C_{ij\to kl,I}(N,\{m^2\},\mu^2)\nonumber\\
  & \times\,\Delta_i (N+1,Q^2,\mu^2)\,\Delta_j (N+1,Q^2,\mu^2)\,
    \Delta^{\rm (s)}_{ij\to kl,I}\bigl(Q/(N\mu),\mu^2\bigr)\,,
\end{align}
where we have introduced the hard scale $Q^2=4m_{\rm av}^2$. The soft radiation is 
coherently sensitive to the colour structure of the hard process from which it 
is emitted \cite{Bonciani:1998vc,Contopanagos:1996nh,Kidonakis:1998bk,Kidonakis:1998nf,Botts:1989kf,Kidonakis:1997gm}. 
At threshold, the resulting colour matrices become diagonal to all orders by 
performing the calculation in the $s$-channel colour basis 
\cite{Beneke:2009rj,Kulesza:2008jb,Kulesza:2009kq}. The different contributions
then correspond to different irreducible representations $I$. Correspondingly, 
$\tilde\sigma^{(0)}_{ij\to kl,I}$ in equation~\eqref{eq:resummed-cross} are the 
colour decomposed leading-order (LO) cross sections. The effects from collinear 
radiation are summed into the functions $\Delta_i$ and $\Delta_j$, and the 
wide-angle soft radiation is described by $\Delta^{\rm (s)}_{ij\to kl,I}$. 
The radiative factors can then be written as
\begin{equation}
  \label{eq:NNLL-expa}
  \Delta_i\Delta_j\Delta^{\rm(s)}_{ij\to kl,I}
  \;=\; \exp\Big[L g_1(\alpha_{\rm s}L) + g_2(\alpha_{\rm s}L) 
                   + \alpha_{\rm s}g_3(\alpha_{\rm s}L) + \ldots \Big]\,.
\end{equation}
This exponent contains all the dependence on large logarithms $L=\log N$. 
The leading logarithmic approximation (LL) is represented by the $g_1$ term 
alone, whereas the NLL approximation requires including the $g_2$ 
term in addition. Similarly, the $g_3$ term is needed for the NNLL approximation. 
The customary expressions for the $g_1$ and $g_2$ functions can be found in 
e.g.~\cite{Kulesza:2009kq} and the one for the NNLL $g_3$ function in 
e.g.~\cite{Beenakker:2011sf}. The matching coefficients $C_{ij\to kl,I}$ 
in~\eqref{eq:resummed-cross} collect non-logarithmic terms as well as 
logarithmic terms of non-soft origin in the Mellin moments of the higher-order 
contributions. At NNLL accuracy the coefficients  $C_{ij\to kl,I}$ factorise into a part that 
contains the Coulomb corrections ${\cal C}_{ij\to kl,I}^{\rm Coul}$ and a part 
containing hard contributions ${\cal C}_{ij\to kl,I}$~\cite{Beneke:2010da}:
\begin{equation}
C_{ij\to kl,I} \;=\; {\cal C}_{ij\to kl,I}^{\rm Coul} \;\times \;
                  (1+\frac{\alpha_{\rm s}}{\pi}\,{\cal C}_{ij\to kl,I}^{(1)}+\dots)\,.
\label{eq:factCcoeff}
\end{equation}

\subsection{Resummation of Coulomb corrections}

In our previous work~\cite{Beenakker:2014sma,Beenakker:2016gmf} we have 
included the Coulomb corrections up to ${\cal O}(\alphas^2)$, i.e. 
$\displaystyle {\cal C}_{ij\to kl,I}^{\rm Coul}
= 1 + \frac{\alpha_{\rm s}}{\pi}\,{\cal C}_{ij\to kl,I}^{\rm Coul,(1)}
    +\frac{\alpha_{\rm s}^2}{\pi^2}\,{\cal C}_{ij\to kl,I}^{\rm Coul,(2)}$\,. 
Here we perform an additional resummation of Coulomb corrections by employing 
the Coulomb Green's function obtained in the NRQCD framework. This results in 
the following expression for the cross section:
\begin{align}
\tilde\sigma^{\mathrm{(res,Coul)}}_{ij \to kl}\bigl(N,\{m^2\},\mu^2\bigr)  
&\;=\; \sum_I \Delta_i\Delta_j\Delta^{\mathrm{(s)}}_{ij\to kl,I}(N+1,Q^2,\mu^2)
       \left(1+\frac{\alphas}{\pi}\,\mathcal{C}_{ij\to kl,I}^{(1)}+\ldots\right)
       \notag\\
\qquad&\hphantom{\;=\ \ }
    \times\,\int_0^1\totd\hat{\rho}\;\hat{\rho}^{N-1}\,
    \sigma^{(0)}_{ij\to kl,I}(\hat{\rho})\,
    \frac{\operatorname{Im}G(0,0;E)}{\operatorname{Im}G^{\mathrm{free}}(0,0;E)}\,,
\label{coulsoftmaster}
\end{align}
with $G(0,0;E)$ being the Green's function at the origin and 
$E = \sqrt{\hat{s}}-2\mav \approx \mav\beta^2 = \mav(1-\hat{\rho})\,$ being the
energy relative to the production threshold. The free Green's function 
$G^{\mathrm{free}}(0,0;E)$ corresponds to freely propagating particles with no 
Coulomb interactions present. Note that in our approach we use the approximation
that the produced sparticles are stable, i.e.~we use the zero-width 
approximation $\Gamma\to 0$. The effect of a non-zero decay width has been studied in \cite{Falgari:2012sq} at NLL accuracy.  
For a moderate width, $\Gamma/m \lesssim 5\%$, the higher-order soft and Coulomb corrections are 
well described by the limit $\Gamma\to 0$, and the remaining ambiguities due to finite-width effects are 
of similar size as the theoretical uncertainties of the threshold-resummed higher-order calculations.

The Green's function up to a certain accuracy is obtained by solving the 
Schr\"odinger equation with a Coulomb potential up to that particular
accuracy~\cite{Beneke:1999qg,Pineda:2006ri}. Up to NLO the Coulomb potential is
given by the following expression in coordinate space:
\begin{align}
V(\mathbf{r}) \;=\; -\,\frac{\alphas(\mu_C) \mathscr{D}_{ij\to kl,I}}{|\mathbf{r}|}
                    \left\{1+\frac{\alphas(\mu_C)}{4\pi}\,
                    \Bigl(8\pi b_0\bigl[\,\log\bigl(\mu_C|\mathbf{r}|\bigr)
                    +\gamma_E\,\bigr]+a_1\Bigr)\right\},\label{coulpotential}
\end{align}
with $\gamma_E = 0.57721...$ the Euler-Mascheroni constant, 
$b_0=(11\,C_A-4\,T_Fn_f)/(12\pi)$ the leading coefficient of the $\beta$-function
for the running of $\alphas$, $a_1 = (31\,C_A-20\,T_F n_f)/9$, and $\mu_C$ 
the Coulomb scale which will be specified in section~\ref{s:numerics}. 
In these coefficients $C_A=N_c$ is the number of colours, $T_F=1/2$ is the 
Dynkin index of the fundamental representation of $SU(N_c)$, and $n_f = 5$ is 
the number of light quark flavours. The colour factor 
$\mathscr{D}_{ij\to kl,I}$ of the QCD Coulomb potential, with the index $I$ 
denoting the colour configuration of the final state, is related to the colour 
operators $T^a_k,T^a_l$ of the final-state particles. It can be expressed in
terms of the quadratic Casimir invariants $C_2(R_I)$, $C_k$ and $C_l$ of the 
particle pair in the colour representation $R_I$ and its constituent particles:
\begin{align}\label{eq:colfac}
\mathscr{D}_{ij\to kl,I} = -\,T^a_k\cdot T^a_l 
\,=\, \frac{1}{2}\left[(T^a_k)^2+(T^a_l)^2-(T^a_k+T^a_l)^2\right] 
\,=\, \frac{1}{2}\bigl(C_k+C_l-C_2(R_I)\bigr)\,.
\end{align}
The general solution to the Schr\"odinger equation for two particles with unequal
masses can be written in the following compact form 
\cite{Beneke:1999qg,Pineda:2006ri,Kiyo:2008bv,Kauth:2011vg,Kauth:2011bz}:
\begin{align}
G(0,0;E+i\Gamma) 
\;=\;  G^{\mathrm{free}}(0,0;E+i\Gamma)
       + \frac{\alphas \mred^2\mathscr{D}_{ij\to kl,I}}{\pi}
         \left[\,g_\mathrm{LO}+\frac{\alphas}{4\pi}\,g_\mathrm{NLO}+\ldots\,\right],
\label{coulgreensfunc}
\end{align}
with $\Gamma = (\Gamma_k+\Gamma_l)/2$ the average decay width of the two 
particles, 
\begin{align}
G^{\mathrm{free}}(0,0;E+i\Gamma) =  i\,\frac{v\mred^2}{\pi}
\label{coulgreensfuncfree}
\end{align}
the free Green's function, and
\begin{align}
g_\mathrm{LO}  &=\; L_v-\psi^{(0)},\notag\\
g_\mathrm{NLO} &=\; 4\pi b_0\bigg[L_v^2-2L_v\left(\psi^{(0)}-\kappa\psi^{(1)}\right)
                   +\kappa\psi^{(2)}+\left(\psi^{(0)}\right)^2-3\psi^{(1)}
                   -2\kappa\psi^{(0)}\psi^{(1)}\notag\\[1mm]
   	     &\hphantom{=\; 4\pi b_0\bigg[}
                   +\,4\,{}_4F_3(1,1,1,1;2,2,1-\kappa;1)\Big]
                   +a_1\Big[L_v-\psi^{(0)}+\kappa\psi^{(1)}\Big]\,.
\label{coulomblonlo}
\end{align}
Here $\psi^{(n)}$ is evaluated with argument $(1-\kappa)$, with 
$\psi^{(n)}(z) = \totd^n/\totd z^n\,\psi^{(0)}(z)$ denoting the $n$-th 
derivative of the digamma function 
$\psi^{(0)}(z) = \gamma_E+\totd/\totd z\log\Gamma(z)$. Furthermore
\begin{align}
\kappa =  i\,\frac{\alphas\mathscr{D}_{ij\to kl,I}}{2v},
\quad L_v = \log\left(\frac{i\mu_C}{4\mred v}\right)
\end{align}
in terms of the non-relativistic velocity
\begin{align}
v = \sqrt{\frac{E+i\Gamma}{2\mred}}\,.
\end{align}
The function ${}_4F_3(1,1,1,1;2,2,1-\kappa;1)$ in \refeq{coulomblonlo} is the 
generalised hypergeometric function\footnote{The efficient numerical evaluation of the 
generalised hypergeometric function is discussed in \cite{Kauth:2011vg}.} and the masses of the 
final-state particles enter the solution in the form of the reduced mass 
$\mred = m_k m_l/(m_k+m_l)$. Setting $\Gamma = 0$ and rewriting 
$E = 2\mred v^2 = \sqrt{\hat{s}}-2\mav$, the non-relativistic velocity $v$ can 
be related to the threshold variable $\beta$ through its definition 
$\beta = \sqrt{1-4\mav^2/\hat{s}}$\,, resulting in 
$v \approx \beta\sqrt{\mav/(2\mred)}\,$ if we neglect threshold-suppressed terms
of $\mathcal{O}(\beta^4)$.

The partonic leading-order cross section $\sigma^{(0)}_{ij\to kl,I}$ splits up 
close to threshold into a $\beta$-dependent part corresponding to the free 
Green's function and a $\beta$-independent hard function 
$\mathcal{H}^{(0)}_{ij\to kl,I}$:
\begin{align}
\sigma^{(0)}_{ij\to kl,I}(\beta) \stackrel{\hat{s}\to 4\mav^2}{\approx} 
\operatorname{Im}G^{\mathrm{free}}(0,0;E)\,\mathcal{H}^{(0)}_{ij\to kl,I} 
\;\approx\; \frac{\mred^2}{\pi}\sqrt{\frac{\mav}{2\mred}}\,\beta\,
            \mathcal{H}^{(0)}_{ij\to kl,I}\,.\label{partinthres}
\end{align}
The proper incorporation of the Coulomb corrections is then achieved by a 
rescaling of the leading-order cross section by the factor 
\cite{Hagiwara:2009hq}:
\begin{align}
\frac{\operatorname{Im}G(0,0;E)}
     {\operatorname{Im}G^{\mathrm{free}}(0,0;E)}\label{coulprescmult}
\end{align}
for every colour channel $I$ separately, giving rise to \refeq{coulsoftmaster}. 
To further increase the accuracy, we supplement the Green's function at NLO 
with the $\log \beta$ corrections originating from two-loop non-Coulombic terms
in the NRQCD potential \cite{Beneke:1999qg,Pineda:2006ri,Beneke:2009ye}. 
This is achieved by multiplying the free Green's function part and the LO 
Coulomb contribution $g_\mathrm{LO}$ in \refeq{coulgreensfunc} by the $\Delta_{\mathrm{nC}}$ factor  presented in \cite{Beneke:2009ye,Beneke:2013opa, Beneke:2014wda}.\footnote{It has been pointed out~\cite{Baernreuther:2013caa, Beneke:2014wda, PSRtalk} that the expression for $\Delta_{\mathrm{nC}}$ in~\cite{Beneke:2009ye,Beneke:2013opa, Beneke:2014wda} is not complete. However, the missing terms contribute to the final result only at the per-mille level. We thank C.\ Schwinn for sharing this observation with us.}

The Green's functions presented in this section and in particular the 
near-threshold expansion \refeq{partinthres} and multiplication prescription 
\refeq{coulprescmult} are only valid for a final state produced with zero 
orbital angular momentum, corresponding to an $s$-wave state. 
Higher partial-wave contributions, such as $p$-wave states, require a 
modification of the Green's function (see e.g.\ \cite{Falgari:2012hx} and 
references therein). Whereas cross sections for $s$-wave states are proportional
to $\beta$ close to threshold, $p$-wave states experience an additional 
phase-space suppression factor of $\beta^2$ relative to the $s$-wave states. 
For the analysis of squark and gluino production at NNLL accuracy presented 
here, the $p$-wave states only give a negligible contribution. They are thus treated 
at NLL accuracy which does not require the inclusion of the Coulomb terms.

\subsection{Inclusion of bound-state contributions}

A very interesting feature of the Coulomb interactions is the possibility of 
bound-state formation at energies below the production threshold of the 
particles. The Green's function then develops single poles of the form 
\cite{Beneke:2005hg}
\begin{align}
G(E+i\Gamma) \;\approx\; \sum_{n=1}^\infty \frac{|\psi_n(0)|^2}{E_n-E-i\Gamma}\,,
\end{align}
where $|\psi_n(0)|^2$ corresponds to the wave function at the origin of the 
bound-state system in the $n^{\rm{th}}$ level with bound-state energy 
$E_n < 0$. In the narrow-width approximation $\Gamma \to 0$, the imaginary part
of the Green's function turns into a sum of sharp $\delta$-peaks:
\begin{align}
\operatorname{Im}G(E+i\Gamma) 
\;=\; \sum_{n=1}^\infty\,|\psi_n(0)|^2\,\frac{\Gamma}{(E_n-E)^2+\Gamma^2} 
\;\stackrel{\Gamma\to 0}{\longrightarrow}\; 
\pi \sum_{n=1}^\infty\,|\psi_n(0)|^2\,\delta(E-E_n)\,.
\end{align}
Only taking into account the LO Coulomb potential, the bound-state energies can
 be computed by analysing \refeq{coulgreensfunc}: the digamma function 
$\psi^{(0)}(1-\kappa)$ is singular for $\kappa = n$, with $n \ge 1$ a positive 
integer. Rewriting $\kappa$ into $E$ gives:
\begin{align}
E_n \;=\; -\,\frac{\alphas^2\mred\mathscr{D}_{ij\to kl,I}^2}{2n^2}\,.
\label{bsenergies}
\end{align}
The residues of $\operatorname{Im}G(E)$ at the bound-state energies precisely
 give the value of the wave function at the origin, i.e.
\begin{align}
|\psi_n(0)|^2 
\;=\; \frac{1}{\pi}\left(\frac{\alphas\mred\mathscr{D}_{ij\to kl,I}}{n}\right)^3.
\end{align}
Therefore, the imaginary part of the bound-state Green's function for a LO 
Coulomb potential becomes
\begin{align}
\operatorname{Im}G^{\mathrm{BS}}(E) 
\;=\; 
\sum_{n=1}^\infty\delta(E-E_n)\left(\frac{\alphas\mred\mathscr{D}_{ij\to kl,I}}{n}
                           \right)^3\quad\text{for}\quad E<0\,.
\label{greenbound}
\end{align}

In this work we consider  the energy levels and residues of a LO Coulomb 
potential\footnote{Bound-state energies and residues for higher-order corrections to the Coulomb potential have been calculated in \cite{Beneke:2005hg}.} 
and suppress the interplay with the soft-gluon corrections, arriving at
\begin{align}
\sigma^{\mathrm{(Coul+BS)}}_{h_1 h_2\to kl}(\rho) 
\;=\; 
\sum_{i,j}\sum_I\int_0^1\totd\tau\,\mathcal{L}_{ij}(\tau)\,
\hat{\sigma}^{\mathrm{(0)}}_{ij\to kl,I}\left(\rho/\tau\right)
\frac{\operatorname{Im}G(0,0;E)}{\operatorname{Im}G^{\mathrm{free}}(0,0;E)}\,,
\end{align}
with the luminosity function
\begin{align}
\mathcal{L}_{ij}(\tau) = \int_\tau^1\frac{\totd x}{x}\,f_{i/h_1}(x,\mu^2)\,
                         f_{j/h_2}\left(\tau/x,\mu^2\right)\,.
\end{align}
While the integration range $\tau\in[\rho,1]$ corresponds to the region above 
threshold where the definitions of \refeqs{coulgreensfunc} and 
\eqref{coulgreensfuncfree} can be used, we are now interested in the 
integration range $\tau\in[0,\rho]$ that denotes the region below threshold. 
Firstly, we apply \refeq{partinthres} to rewrite the cross section close to 
threshold and subsequently replace $\operatorname{Im}G(0,0;E)$ with 
\refeq{greenbound} to obtain
\begin{align}
\sigma^{\mathrm{BS}}_{h_1 h_2\to kl}(\rho) 
\;=\; \sum_{i,j}\sum_I\int_0^{\rho}\totd\tau\,\mathcal{L}_{ij}(\tau)\,
      \mathcal{H}^{(0)}_{ij\to kl,I}\sum_{n=1}^\infty\delta(E-E_n)
      \left(\frac{\alphas\mred\mathscr{D}_{ij\to kl,I}}{n}\right)^3.
\end{align}
Rewriting  the delta function 
\begin{align}
\delta(E-E_n) \;=\; \delta\left(\sqrt{\tau S}-2\mav-E_n\right) 
\;=\; \delta(\tau-\tau_n)\,\frac{2(E_n+2\mav)}{S}
\end{align}
with $\tau_n = (E_n+2\mav)^2/S$, the integration over $\tau$ can be performed 
trivially. Additionally, we improve the description of the partonic cross 
section near threshold by replacing the flux factor at threshold 
$1/\hat{s}_{\mathrm{thr}} \approx 1/(4\mav^2)$ with the exact flux factor 
$1/\hat{s}$, resulting in
\begin{align}
\sigma^{\mathrm{BS}}_{h_1 h_2\to kl}(\rho) 
&\;=\; \sum_{i,j}\sum_I\sum_{n=1}^\infty\mathcal{L}_{ij}(\tau_n)\,
       \frac{4\mav^2}{\tau_nS}\,\mathcal{H}^{(0)}_{ij\to kl,I}\,
       \frac{2(E_n+2\mav)}{S}\left(\frac{\alphas\mred\mathscr{D}_{ij\to kl,I}}{n}
                             \right)^3\notag\\
&\;=\; \sum_{i,j}\sum_I\sum_{n=1}^\infty\mathcal{L}_{ij}(\tau_n)\,
       \mathcal{H}^{(0)}_{ij\to kl,I}\,\frac{8\mav^2}{(E_n+2\mav)S}
       \left(\frac{\alphas\mathscr{D}_{ij\to kl,I}}{n}\right)^3.
\label{finalbsformimp}
\end{align}
It should be noted that in the sum over the colour channels $I$, only channels 
corresponding to an attractive Coulomb potential, 
i.e.~$\mathscr{D}_{ij\to kl,I} > 0$, are summed over. 

The bound-state contributions below threshold are then added to the NNLL 
predictions above threshold including Coulomb resummation, matched in the 
Mellin-space formalism to the approximated 
next-to-next-to-leading order cross section NNLO$_\mathrm{Approx}$. 
The latter is constructed by adding the near-threshold approximation of the 
NNLO correction~\cite{Beneke:2009ye} to the full NLO 
result~\cite{Beenakker:1996ch}. As the final result, we thus obtain
\begin{align}
\label{eq:matching}
&\sigma^{\rm (NNLL~matched,\,Coul+BS)}_{h_1 h_2 \to kl}\bigl(\rho, \{m^2\},\mu^2\bigr) 
 \;=\; \sigma^{\mathrm{BS}}_{h_1 h_2\to kl}(\rho) 
       \,+\, \sigma^{\rm (NNLO_{Approx})}_{h_1 h_2 \to kl}\bigl(\rho,\{m^2\},\mu^2\bigr)
       \nonumber\\[2mm]
&\hspace*{2.5ex}
 +\, \sum_{i,j}\,\int_\mathrm{CT}\,\frac{dN}{2\pi i}\;\rho^{-N}\,
     \tilde f_{i/h_1}(N+1,\mu^2)\,\tilde f_{j/h_{2}}(N+1,\mu^2) \nonumber\\[2mm]
&\hspace*{5.5ex}\times\,
 \left[\,\tilde\sigma^{\rm(res,\,NNLL,\, Coul)}_{ij\to kl}\bigl(N,\{m^2\},\mu^2\bigr)
         \,-\, \tilde\sigma^{\rm(res,\,NNLL,\,Coul)}_{ij\to kl}\bigl(N,\{m^2\},\mu^2
               \bigr){\left.\right|}_{\scriptscriptstyle{\rm (NNLO)}}\,\right]\,.
\end{align}
To evaluate the inverse Mellin transform in~\eqref{eq:matching} we adopt the ``minimal prescription'' of reference~\cite{Catani:1996yz} for the integration contour CT. 

\section{Numerical results}\label{s:numerics}
In this section we present numerical results for the NNLL resummed cross sections matched to the approximated NNLO results for pair production of squarks and gluinos at the LHC with $\sqrt S = 13$ TeV.

As already mentioned, all flavours of final-state squarks are included and summed over, except for stops, which are treated separately because of the large mixing effects and the mass splitting in the stop sector. We sum over squarks with both chiralities ($\tilde{q}_{L}$ and~$\tilde{q}_{R}$), which are taken as mass degenerate. All light-flavour squarks are also assumed to be mass degenerate. The QCD coupling $\alpha_{\rm s}$ and the parton distribution functions at NLO and NNLO are defined in the $\overline{\rm MS}$ scheme with five active flavours, and a top-quark mass of $m_t=173.2$~GeV~\cite{Agashe:2014kda} is used. At NLO and beyond, the stop cross section depends not only on the stop mass, but also on the masses of the gluino and the other squark flavours, and on the stop mixing angle, all of which enter through loop contributions. We present the stop cross sections in a simplified SUSY scenario, where only the lighter stop mass eigenstate and the gluino are accessible at the LHC, while all other squark flavours are decoupled. In the stop cross sections presented below we set the gluino mass equal to the stop mass. The dependence of the cross section on the stop mixing angle $\theta_{\tilde{t}}$ is small, with changes of typically less than 5\%; for definiteness we have set $\sin 2\theta_{\tilde{t}}= 0.669$ for the numerical evaluation~\cite{Beenakker:2016gmf}.

As our default choice for NNLL and approximated NNLO calculations, we use the NNLO PDF4LHC15\_mc set of parton distribution functions (pdfs)~\cite{Butterworth:2015oua}. The NLO and NLL results presented for reference are obtained using the NLO PDF4LHC15\_mc pdf sets~\cite{Butterworth:2015oua}. Both at NLO and NNLO the value of $\alpha_{\rm s}(M_{Z}) = 0.118$ is used. In order to use standard parametrizations of pdfs in $x$-space, we employ the method introduced in reference~\cite{Kulesza:2002rh}. We note that the impact of threshold-improved parton distributions on the squark and gluino cross-section predictions has been studied in \cite{Beenakker:2015rna} at NLL accuracy. 

In the numerical calculations the renormalisation and factorisation scales are taken to be equal $\mu=\mu_R=\mu_F=\mav$.   The relevant (Bohr) scale for bound-state effects is defined as twice the Bohr radius $r_B$ of the corresponding bound states~\cite{Hagiwara:2008df,Kiyo:2008bv,Hagiwara:2009hq,Beneke:2010da,Falgari:2012hx}. For a LO Coulomb potential it amounts to 
\begin{align}
\mu^{[I]}_B = \frac{2}{r_B} \;=\; 2m_{\mathrm{red}}\,\big|\mathscr{D}_{ij\to kl,I}\big|\,\alphas\left(\mu^{[I]}_B\right),\label{bohrscale}
\end{align}
where $\mathscr{D}_{ij\to kl,I}$ denotes the colour factor of the Coulomb potential defined in \refeq{eq:colfac}, and $\mred = m_k m_l/(m_k+m_l)$ is the reduced mass of the final-state particles. 
The QCD coupling $\alphas$ is evaluated at the Bohr scale itself, requiring \refeq{bohrscale} to be solved iteratively. It should be noted that the Bohr scale depends on the colour channel. \Refeq{bohrscale} defines the scale that is used for the bound-state contributions, \refeq{finalbsformimp}. We note that in \refeq{finalbsformimp}, the $\alphas$ factors of $\mathcal{H}^{(0)}_{ij\to kl,I}$, stemming from the off-shell interactions, are evaluated at the usual value of the renormalisation scale $\mu_R$ close to the typical hard scale $\sim\mav$ of the process, whereas the $\alphas$ factors originating from the Green's function are evaluated at the Bohr scale as defined in \refeq{bohrscale} for the attractive colour channels with $\mathscr{D}_{ij\to kl,I} > 0$. 

As argued in \cite{Beneke:2010da,Falgari:2012hx}, from momentum power-counting of the Coulomb gluons and by analysing the scale dependence of the terms in \refeq{coulomblonlo}, the natural scale of Coulomb interactions turns out to be
\begin{align}
	\mu_C \sim 4\mred v \;\approx\; 2\beta\sqrt{m_km_l}\label{coulscale1}\,.
\end{align}
This choice of $\mu_C$ minimises the effect of a residual higher-order scale dependence from a truncation of the perturbative expansion of $\alphas$. The Coulomb scale appears both explicitly in the Green's function and implicitly in the factors of $\alphas$ for the Coulomb contributions above threshold in \refeqs{eq:matching} and (\ref{coulsoftmaster}).
Such a scale choice also leads to the inclusion of some of the mixed soft $\log(\beta)$ and Coulomb $1/\beta$ terms such as $\alphas\log(\beta)\,\times\,\alphas/\beta$, which are formally of NLL accuracy \cite{Beneke:2010da}. The scale choice of \refeq{coulscale1} needs to be bounded from below, since, besides hitting the Landau pole when $\beta$ is integrated over, bound-state effects become important in the vicinity of the threshold. Therefore, we use the Bohr scale of \refeq{bohrscale} to bound the Coulomb scale from below. While bound states only arise for an attractive Coulomb potential, $\mathscr{D}_{ij\to kl,I} > 0$, we also use the Bohr scale as a lower bound for a repulsive potential. This is not completely justified, but leads to negligible effects for the total cross section above threshold \cite{Falgari:2012hx}. The Coulomb scale is consequently chosen to be
\begin{align}
	\mu_C = \max\left\{\mu_B^{[I]},2\beta\sqrt{m_km_l}\right\},\label{coulscale}
\end{align}
which ensures that, as soon as the bound-state effects become relevant, the Bohr scale is used.
We stress that \refeq{bohrscale} is used for the contributions below threshold, while \refeq{coulscale} applies to the contributions above threshold.

In the following discussion we present predictions for the LHC squark and gluino cross sections for a center-of-mass energy of 13~TeV,  at various levels of theoretical accuracy:

\begin{itemize}

\item The NLO cross sections~\cite{Beenakker:1996ch}, denoted as $\sigma^{\rm NLO}$.

\item The NLL cross sections matched to NLO results, based on the calculations presented in \cite{Beenakker:2009ha,Kulesza:2008jb,Kulesza:2009kq}, and corresponding to the output of the \texttt{NLL-fast} numerical package. They are denoted as $\sigma^{\rm NLO+NLL}$.

\item The NNLL matched cross sections including Coulomb corrections up to ${\cal O}(\alphas^2)$, but without bound-state contributions, which are excluded by setting  $\sigma^{\mathrm{BS}}_{h_1 h_2\to kl}=0$. These cross sections are referred to as ``NNLO$_{\rm Approx}$+NNLL'' in the plots. This level of accuracy corresponds to the one discussed in detail in~\cite{Beenakker:2014sma}.  The NNLO$_{\rm Approx}$ +NNLL accuracy, as detailed in Eq.~(\ref{eq:matching}), applies to the $s$-wave channels. The contributions from the $\beta^2$-suppressed $p$-wave channels are taken into account at NLO+NLL accuracy. Both the $s$-wave and the $p$-wave contributions have been convoluted with the PDF4LHC15\_mc  NNLO parton distribution functions. In our previous work~\cite{Beenakker:2014sma}, we have checked that using NNLO pdfs instead of NLO pdfs for the suppressed $p$-wave contributions leads to a negligible modification of the full result.

\item The NNLL matched cross sections including resummed Coulomb corrections, but without bound-state contributions, corresponding to \refeq{eq:matching} with  $\sigma^{\mathrm{BS}}_{h_1 h_2\to kl}=0$. These cross sections are referred to as ``NNLO$_{\rm Approx}$+NNLL+Coul'' in the plots. 

\item The NNLL matched cross sections including resummed Coulomb corrections and bound-state contributions,  corresponding to \refeq{eq:matching}, which are called 
``NNLO$_{\rm Approx}$ +NNLL+Coul+BS'' or ``NNLL-fast'' in the plots. 

\end{itemize}
Apart from the NLO cross sections, which were calculated using the publicly available {\tt PROSPINO} code~\cite{prospino}, all our results were obtained using two independent computer codes.

\medskip

\begin{figure}[h]
	\centering
	\includegraphics[width=0.65\textwidth]{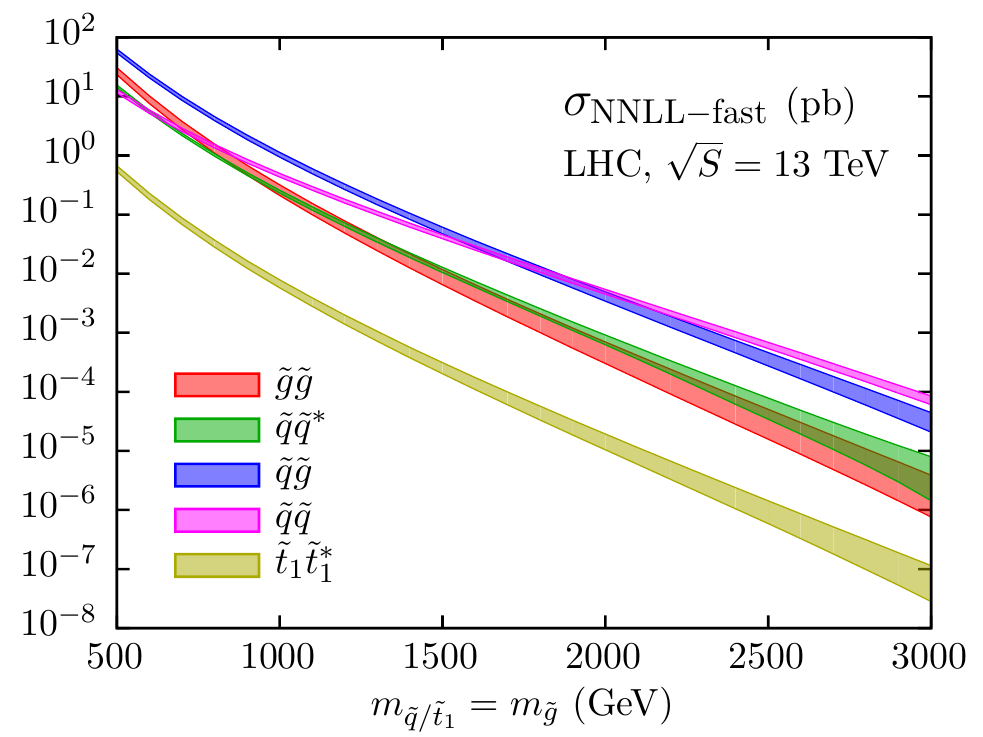}
	\caption{Cross section predictions for squark and gluino production at the LHC  with $\sqrt{S} = 13$\,TeV at NNLO$_\mathrm{Approx}$+NNLL accuracy, including Coulomb resummation and bound state effects. The error bands denote the theoretical uncertainty due to scale variation and the pdf+$\alpha_{\rm s}$ error as described in the text.}
	\label{fig:nnllplusnnlo}
\end{figure}
We begin our discussion of the numerical results by presenting in~\reffig{fig:nnllplusnnlo} our best prediction for the total cross sections at  NNLO$_{\rm Approx}$+NNLL+Coul+BS accuracy 
as a function of the mass of the final-state sparticles, for the special case of equal squark and gluino masses, $m_\squ=m_\glu$. The error bands denote the overall theoretical uncertainty obtained from the variation of the renormalisation, factorisation and Coulomb scales by a factor of two about the corresponding central scales, and from the pdf and $\alpha_{\rm s}$ error as evaluated using the NNLO PDF4LHC15\_mc pdf set~\cite{Butterworth:2015oua}, added linearly to the scale uncertainty. Note that the increase in the overall uncertainty at large sparticle masses is due to the pdf uncertainty, see e.g.\ \cite{Beenakker:2015rna}, and is particularly pronounced for those production channels that are driven by the gluon luminosity. 

\begin{figure}[h]
	\centering
	\includegraphics[width=0.47\textwidth]{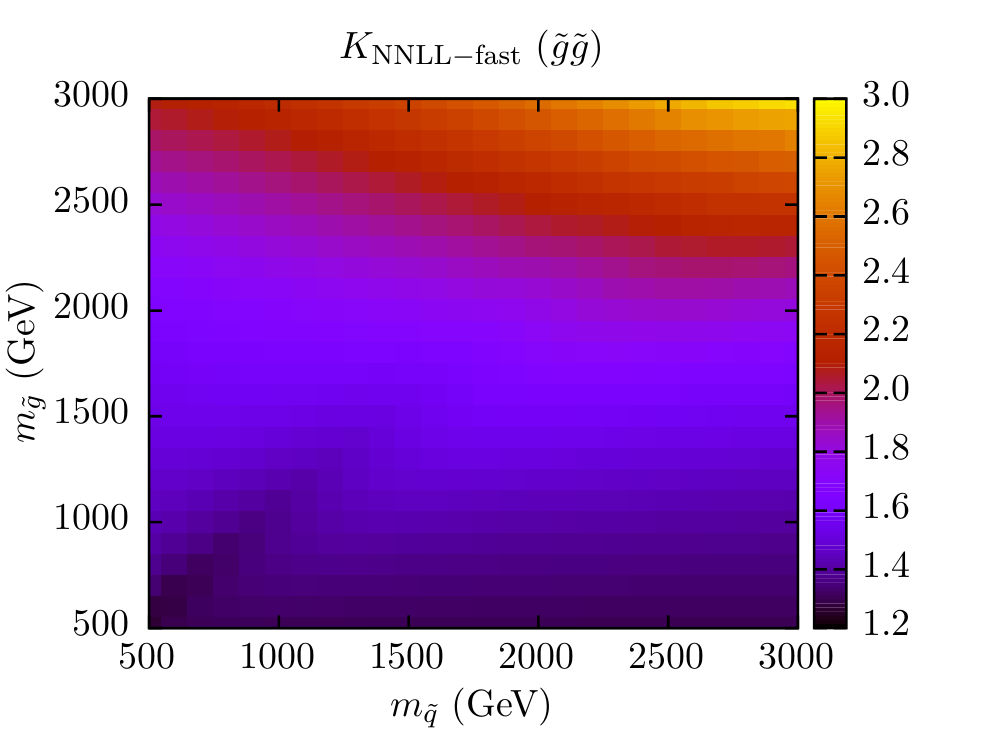}\includegraphics[width=0.47\textwidth]{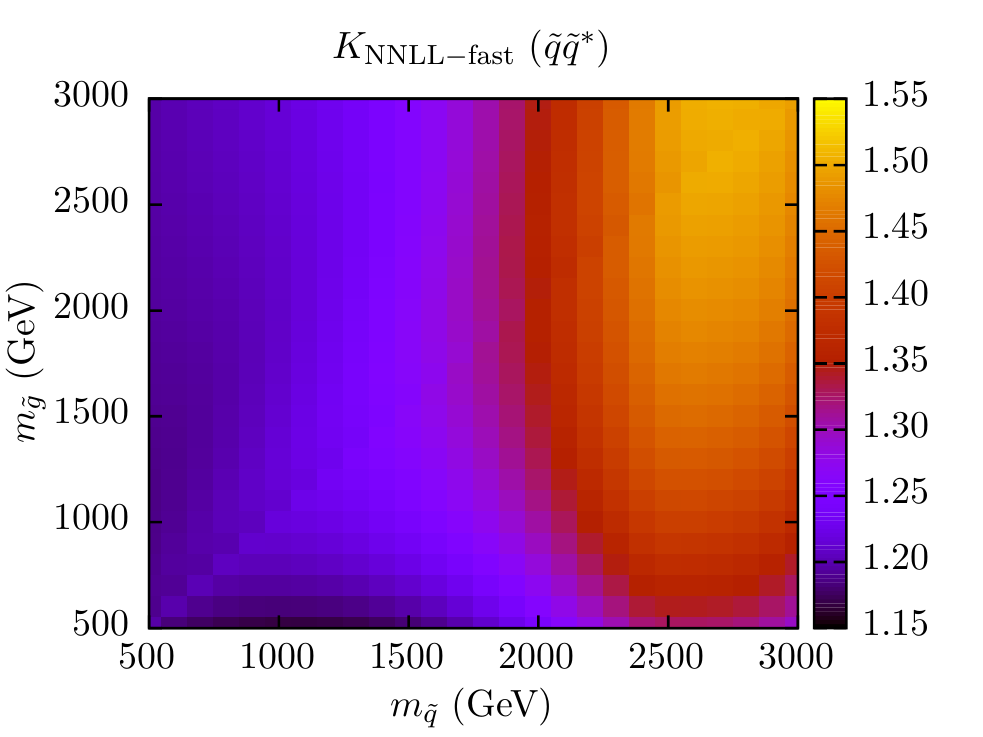}\\
	\includegraphics[width=0.47\textwidth]{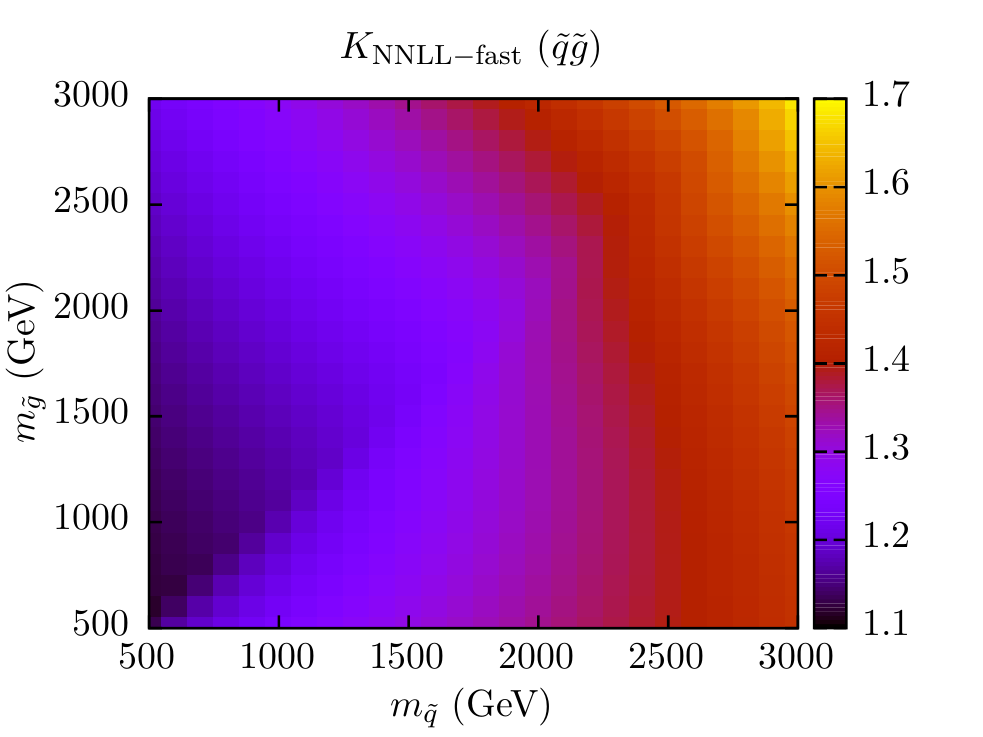}\includegraphics[width=0.47\textwidth]{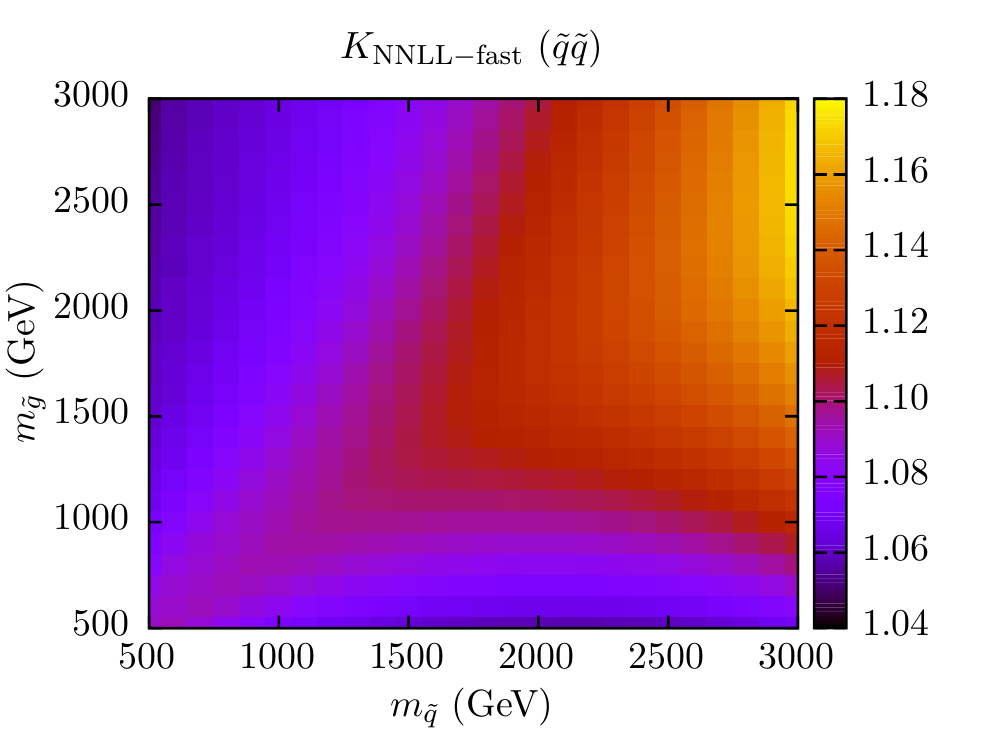}\\
	\includegraphics[width=0.47\textwidth]{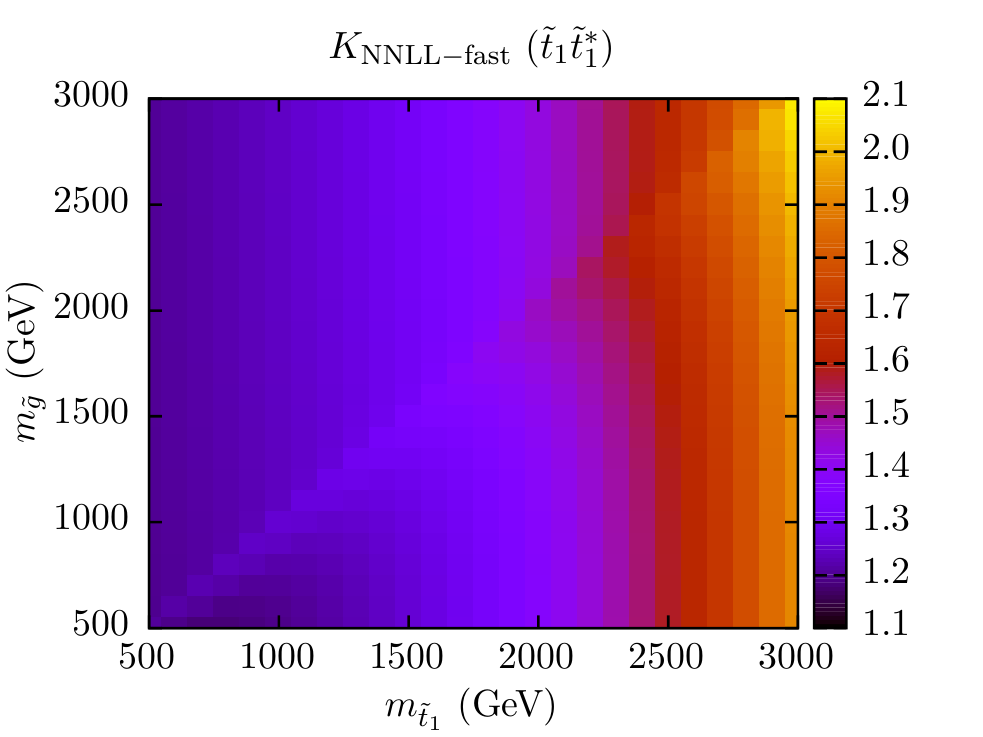}
	\caption{$K$-factors for squark and gluino production at the LHC with $\sqrt{S} = 13$\,TeV at NNLO$_\mathrm{Approx}$+NNLL accuracy, including Coulomb resummation and bound state effects. The $K$-factors are evaluated  with respect to the NLO cross sections and displayed as a function of the squark and gluino masses.}
	\label{fig:k2d}
\end{figure}
In \reffig{fig:k2d} we present the NNLO$_{\rm Approx}$+NNLL+Coul+BS $K$-factors, with respect to the NLO cross sections evaluated with NLO pdfs, as a function of the squark and gluino masses. The $K$-factors range from close to one to up to a factor of three, depending in detail on the production process and the masses of the final state particles. The effect of soft-gluon resummation is most pronounced for processes with initial-state gluons and final-state gluinos, which involve a large colour charge. As expected, the resummation corrections become larger for increasing sparticle masses because of the increasing importance of the threshold region. 

\begin{figure}[h]
	\centering
	\includegraphics[width=0.47\textwidth]{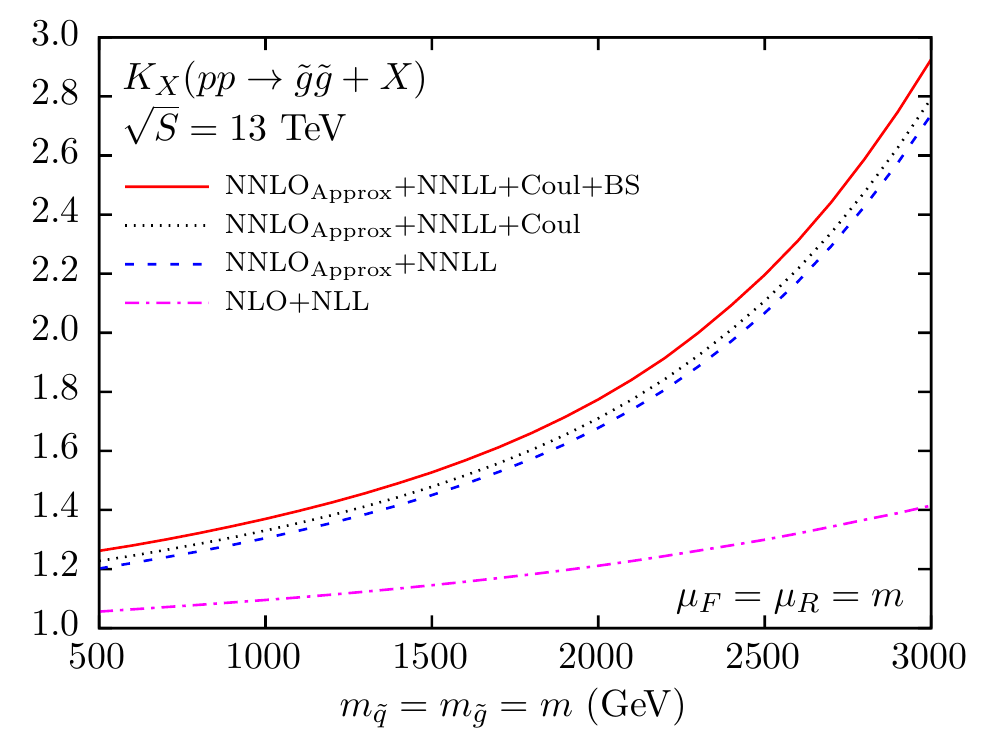}\includegraphics[width=0.47\textwidth]{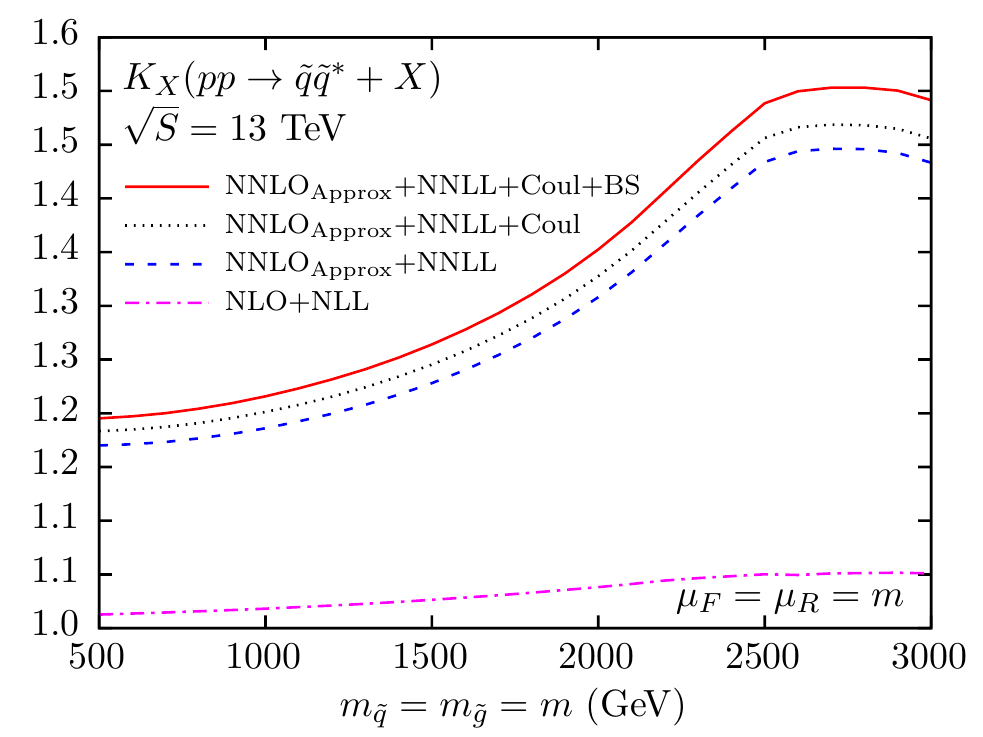}\\
	\includegraphics[width=0.47\textwidth]{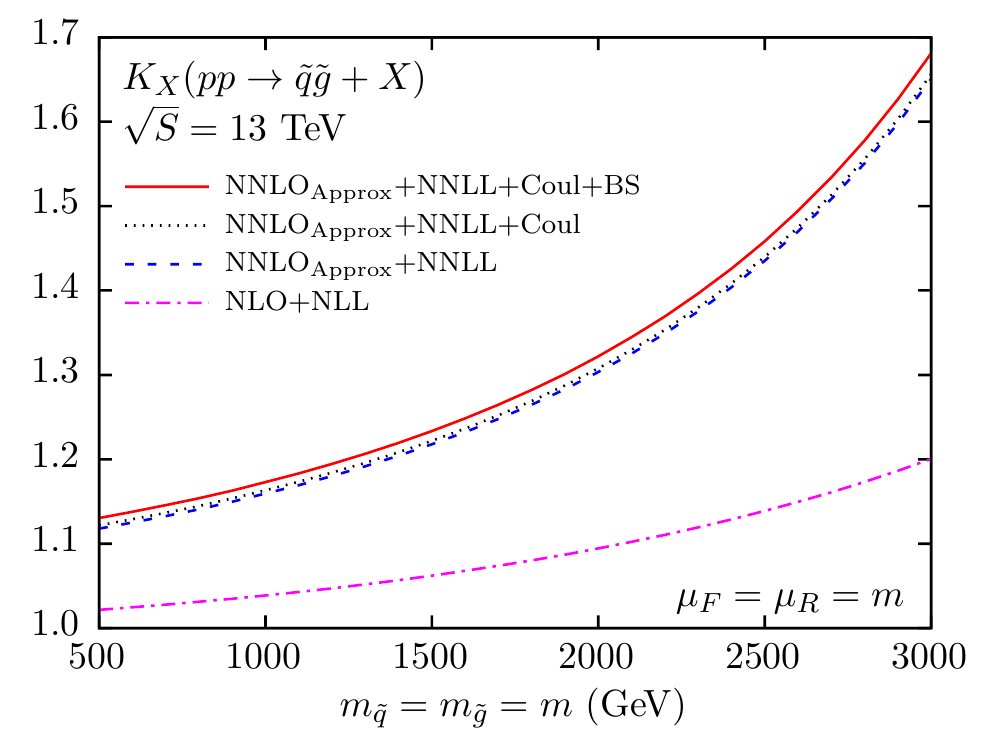}\includegraphics[width=0.47\textwidth]{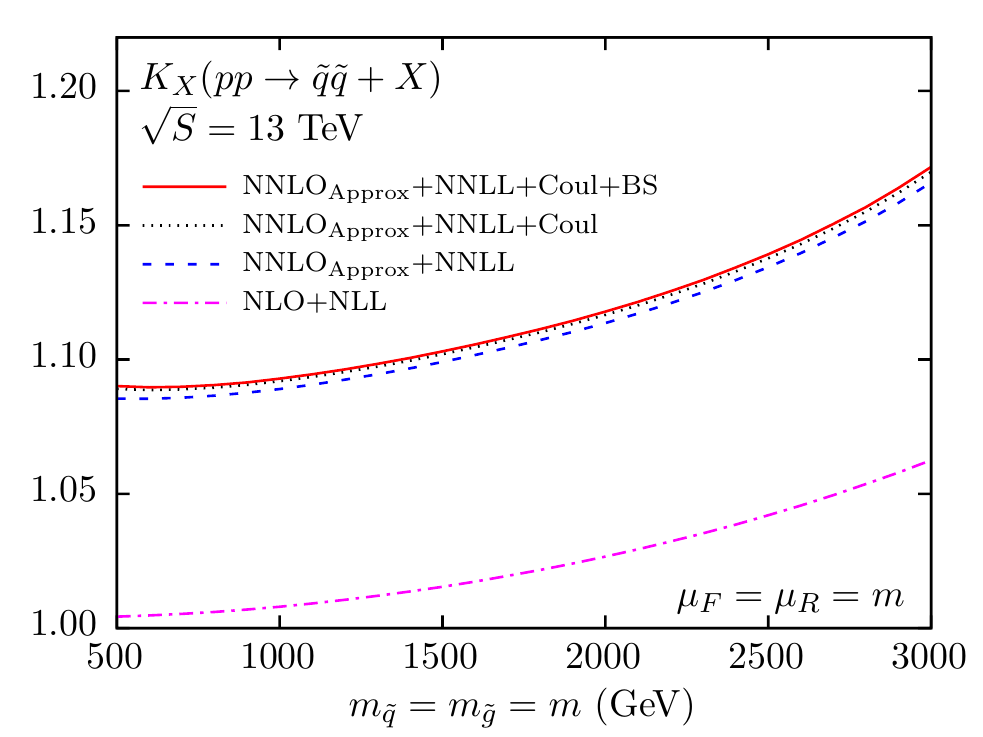}\\
	\includegraphics[width=0.47\textwidth]{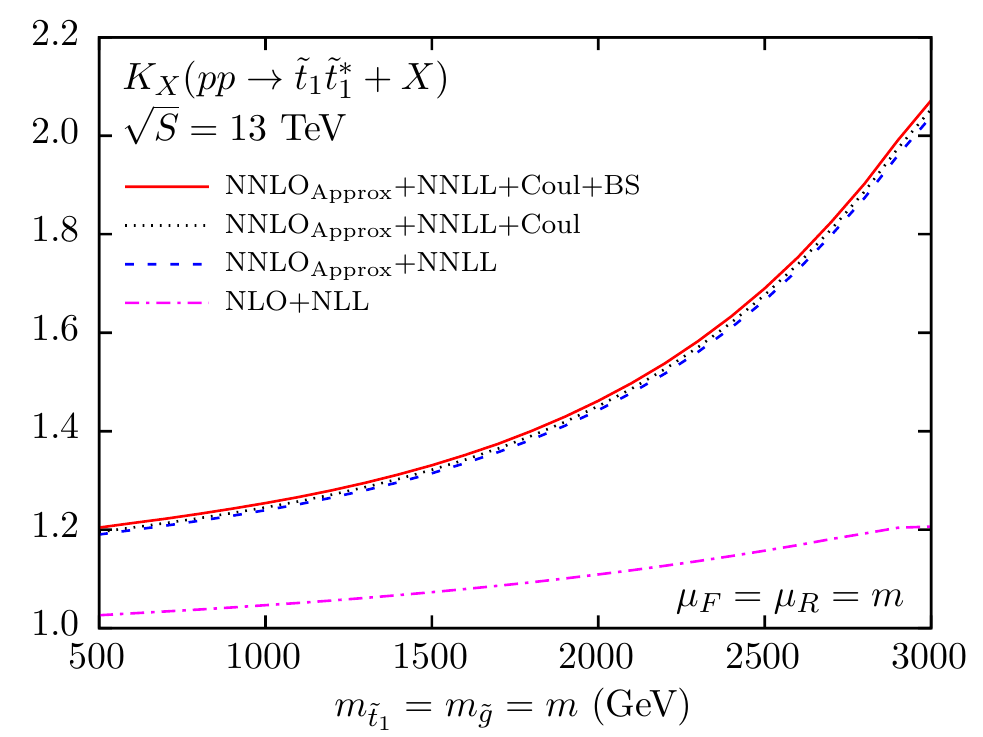}
	\caption{$K$-factors for squark and gluino production at the LHC with $\sqrt{S} = 13$\,TeV at various levels of theoretical accuracy. The $K$-factors are evaluated  with respect to the NLO cross sections and displayed as a function of the common squark and gluino mass.}
	\label{fig:coulbsk}
\end{figure}
A more detailed analysis of the size of the $K$-factors with respect to the NLO prediction is presented in \reffig{fig:coulbsk}, where we compare the $K$-factors obtained at NLO+NLL, NNLO$_{\rm Approx}$+NNLL, NNLO$_{\rm Approx}$+NNLL+Coul, 
and NNLO$_{\rm Approx}$+NNLL+Coul +BS  accuracy. We find that the NNLL corrections are positive and significantly enhance the NLO+NLL cross sections. The importance of the NNLL threshold corrections increases with increasing final-state particle masses, as discussed in \cite{Beenakker:2014sma,Beenakker:2016gmf}. 
The contributions from Coulomb resummation and bound states are also positive and result in a slight further increase of the $K$-factors. 

The squark-antisquark and gluino-pair production cross sections computed with some replicas of the NNLO PDF4LHC15\_mc pdf sets become negative for squark and gluino masses beyond about 2\,TeV.  Following the prescription discussed in \cite{Beenakker:2015rna} we set the corresponding cross sections to zero before evaluating the central cross section prediction from all replicas. The occurrence of negative cross sections is most pronounced for  
squark-antisquark production and leads to a decrease of the corresponding $K$-factor at large squark masses, see upper right panel in \reffig{fig:coulbsk}. The occurrence of negative cross section predictions and the resulting peculiar behaviour of the squark-antisquark $K$-factor are a further indication that the accuracy of cross section predictions for heavy new particles is currently limited by the pdf uncertainty.

\begin{figure}[h]
	\centering
	\includegraphics[width=0.47\textwidth]{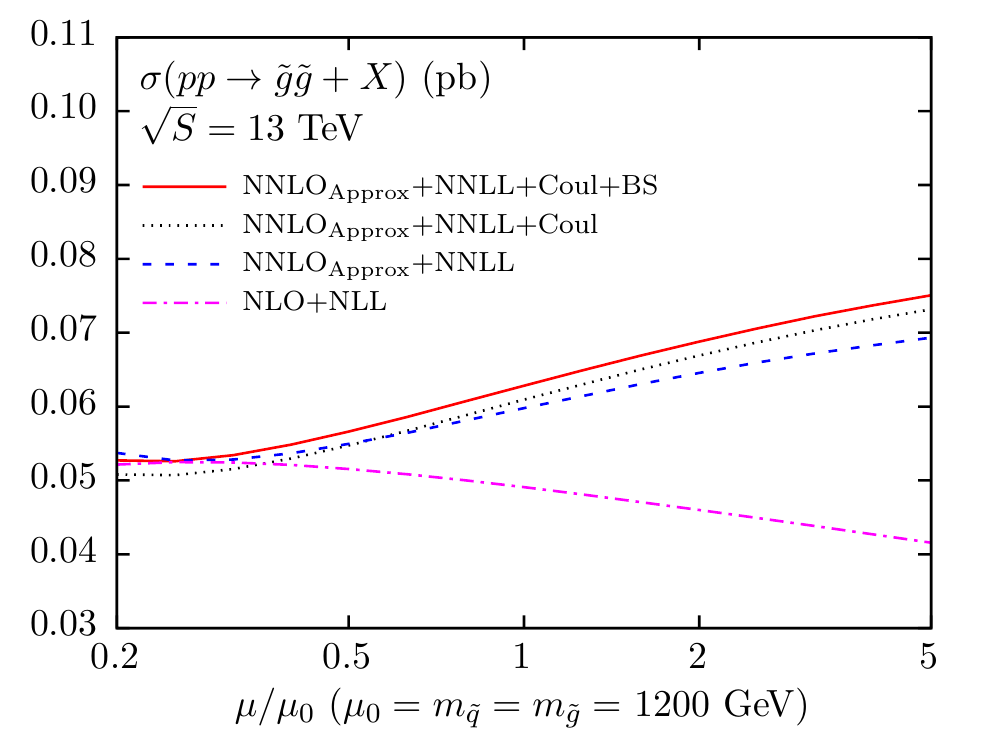}\includegraphics[width=0.47\textwidth]{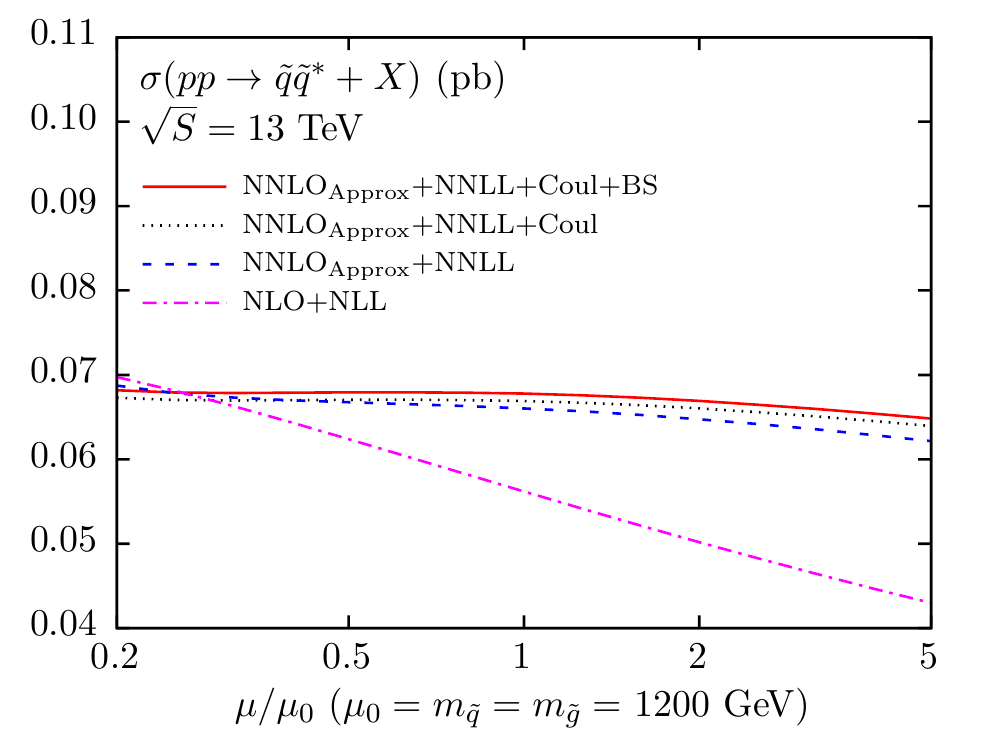}\\
	\includegraphics[width=0.47\textwidth]{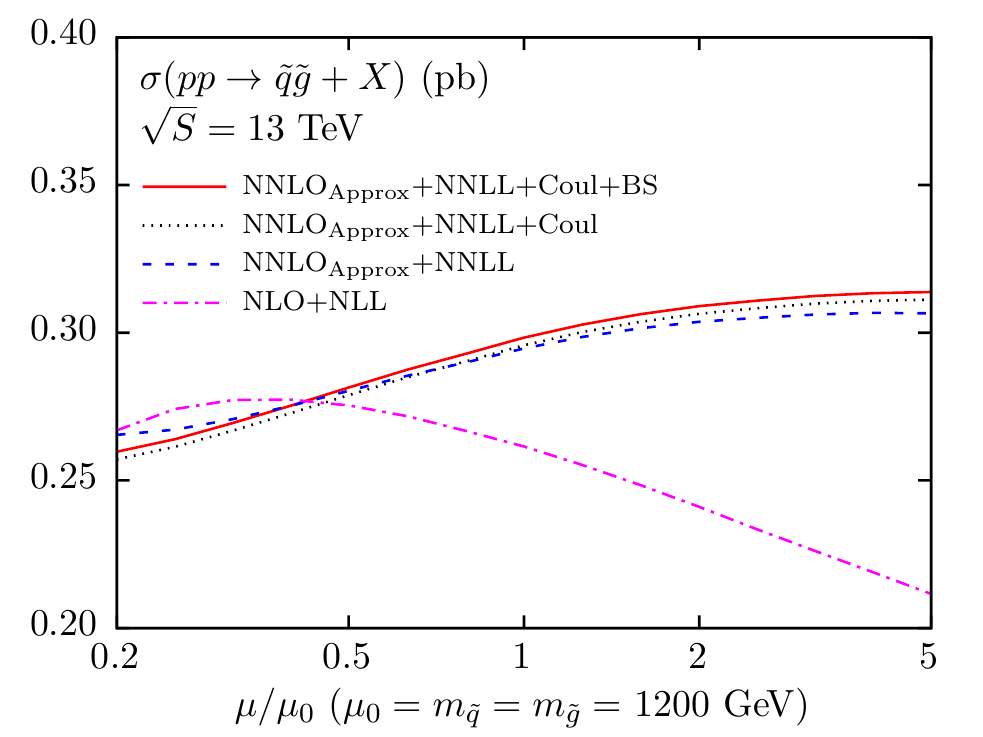}\includegraphics[width=0.47\textwidth]{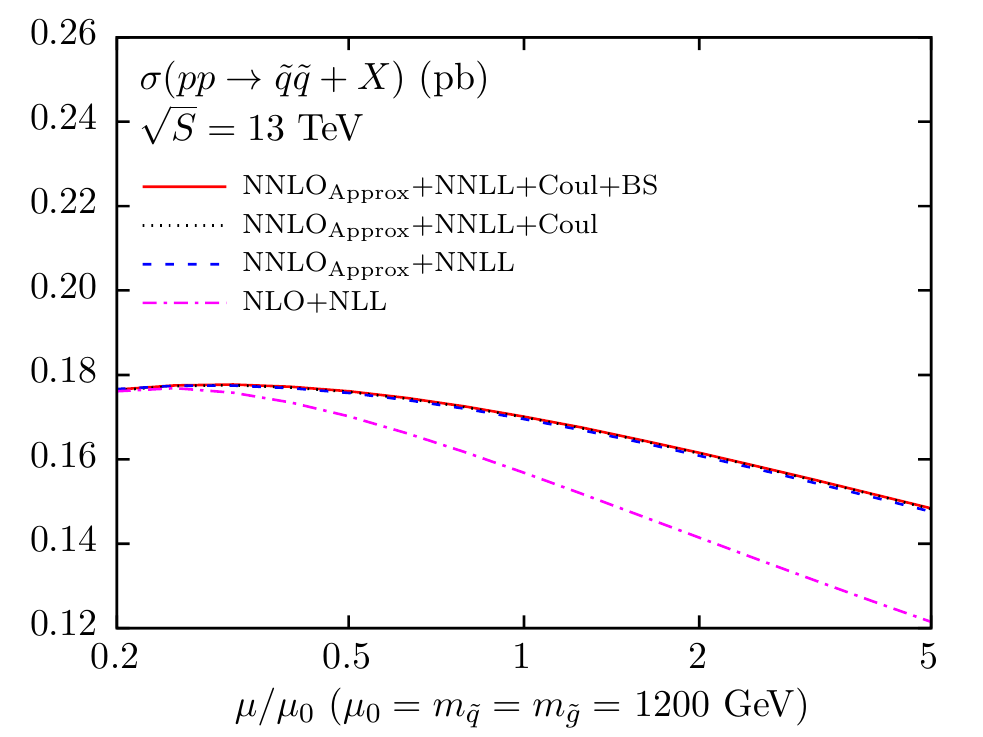}\\
	\includegraphics[width=0.47\textwidth]{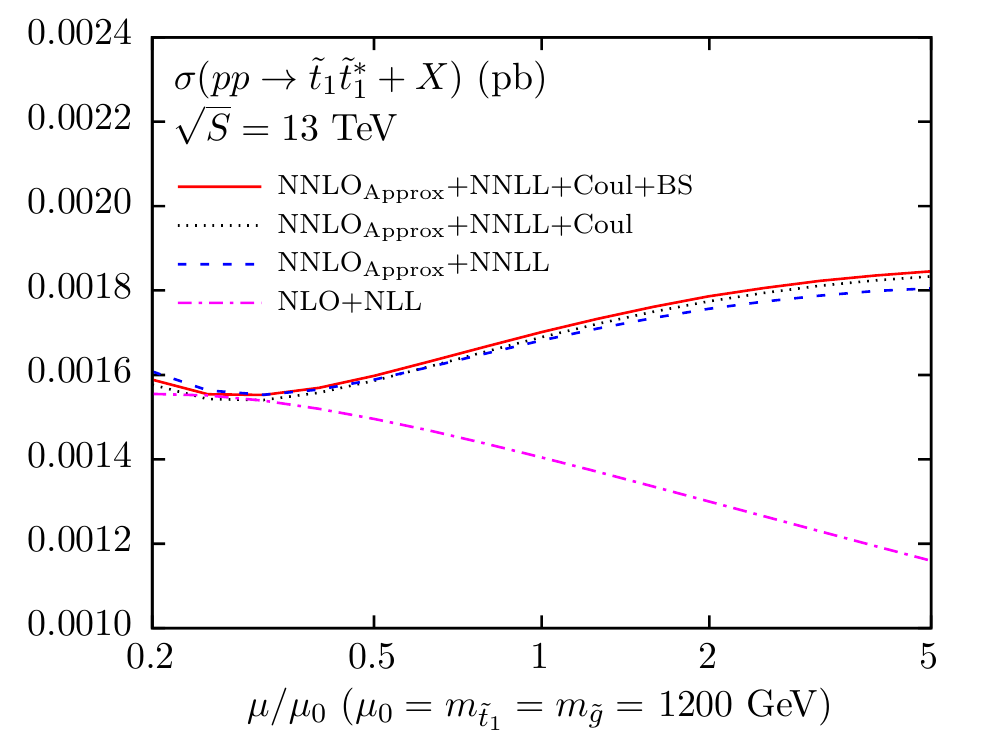}
	\caption{Scale dependence of the squark and gluino cross section at the LHC with $\sqrt{S} = 13$\,TeV for squark and gluino masses of 1.2\,TeV. Shown is the dependence of the cross section on the renormalization and factorization scales, with the Coulomb scale kept fixed, for various levels of theoretical accuracy.}
	\label{fig:coulbsmu}
\end{figure}
We finally investigate the scale dependence of the different squark and gluino production processes, see \reffig{fig:coulbsmu}. The squark and gluino mass are both set to 1.2\,TeV, and we vary the renormalization and factorization scales by a factor of five about the 
central value $\mu_0 = m_{\tilde{q}}= m_{\tilde{g}}$, while the Coulomb scale is fixed to its default value \refeq{coulscale}.\footnote{We note that the matching procedure employed in Eq.~(\ref{eq:matching})  introduces different scales in the Coulomb terms: the resummed higher-order Coulomb corrections are calculated at the Coulomb scale $\mu_C$ while the corrections included in the fixed-order expressions are calculated at the scale $\mu$.} \Reffig{fig:coulbsmu} shows that the scale dependence in general decreases when including NNLL resummation. The NNLL scale dependence of gluino-pair production, however, is not improved compared to NLO+NLL, which can be understood from an intricate interplay of various effects, as discussed in reference~\cite{Beenakker:2014sma}. 
The inclusion of Coulomb resummation and bound-state contributions only has a small impact on the overall scale uncertainty.

\subsection{Comparison with SCET results}
The joint resummation of threshold and Coulomb corrections has first been considered in the framework of SCET~\cite{Beneke:2010da,Falgari:2012hx}. It is instructive to see how well the two independent approaches of threshold resummation agree, and to study the significance of the subleading contributions, which may differ in the two formalisms. 

In this section we compare our results with those of ref.~\cite{Beneke:2014wda}, which we denote ``NNLL (SCET)'', corresponding to the currently best result within the SCET formalism for squark and gluino production. ``NNLL (SCET)'' includes the soft exponentials up to NNLL, one-loop hard-matching coefficients, Coulomb resummation with a NLO Coulomb potential, bound states, and matching to NNLO$_{\mathrm{Approx}}$. We also
show ``NNLL$_{\mathrm{fixed}-C}$ (SCET)'' predictions, which correspond to ``NNLL (SCET)'' with the Coulomb contributions expanded up to $\mathcal{O}(\alphas^2)$.  The results of~\cite{Beneke:2014wda} are only available for the production of light-flavour squarks and gluinos. The results for stop-pair production presented in~\cite{Broggio:2013cia} do not include the resummation of the Coulomb corrections and have been compared to the corresponding results in the Mellin-space formalism in our earlier work~\cite{Beenakker:2016gmf}.

The SCET results have been read off the plots in~\cite{Beneke:2014wda}. For the sake of comparison we present the cross-sections at a collider energy of $\sqrt{S} = 8$ TeV and use the MSTW2008 pdf set, with NLO pdfs for results at NLO and NLO+NLL accuracy, and NNLO pdfs for results at NNLO$_{\mathrm{Approx}}$+NNLL accuracy. We will show our predictions for the NNLO$_\mathrm{Approx}$+NNLL, NNLO$_\mathrm{Approx}$+NNLL+Coul and NNLO$_\mathrm{Approx}$+NNLL+Coul+BS accuracies. Comparing our predictions in the Mellin-space formalism with the SCET formalism, NNLO$_\mathrm{Approx}$ +NNLL+Coul+BS corresponds to NNLL (SCET) and NNLO$_\mathrm{Approx}$+NNLL approximately corresponds to NNLL$_{\mathrm{fixed}-C}$ (SCET). This is not an exact correspondence, as there are the following differences between our predictions and those from \cite{Beneke:2014wda}:
\begin{itemize}
	\item The expressions in the SCET formalism depend on more scales that can be varied independently. In particular the soft scale can be chosen independently, while it is fixed to the value $Q/N$ in the Mellin-space formalism.
	\item The SCET framework allows one to convolute the bound-state contributions with the soft function as described in \cite{Beneke:2011mq}.  Furthermore, the authors of \cite{Beneke:2014wda}  also include bound-state effects from a NLO Coulomb potential, causing a small shift in the bound-state energies and residues, while we only include bound-state effects originating from a LO potential.
\end{itemize}

\begin{figure}[h]
	\centering
	\includegraphics[width=0.47\textwidth]{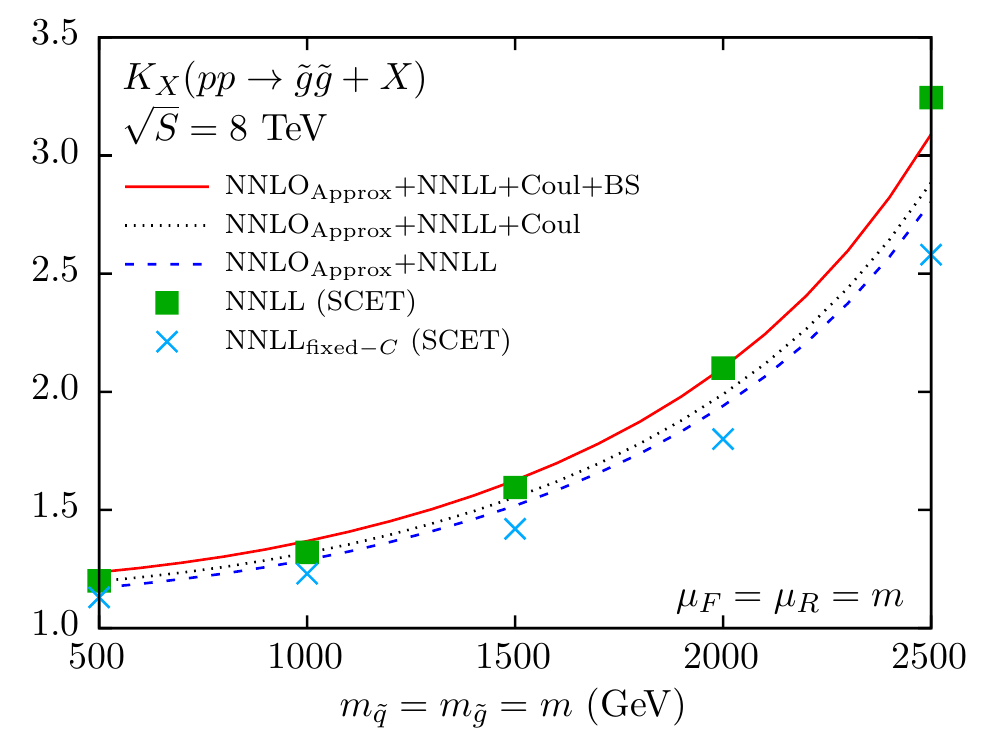}\includegraphics[width=0.47\textwidth]{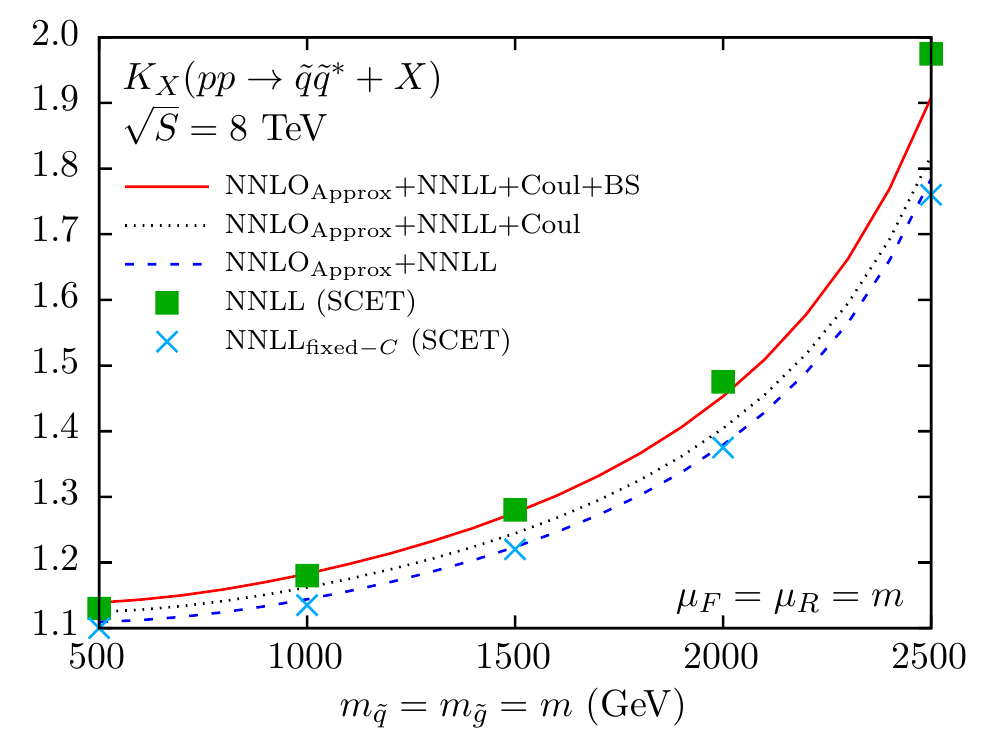}\\
	\includegraphics[width=0.47\textwidth]{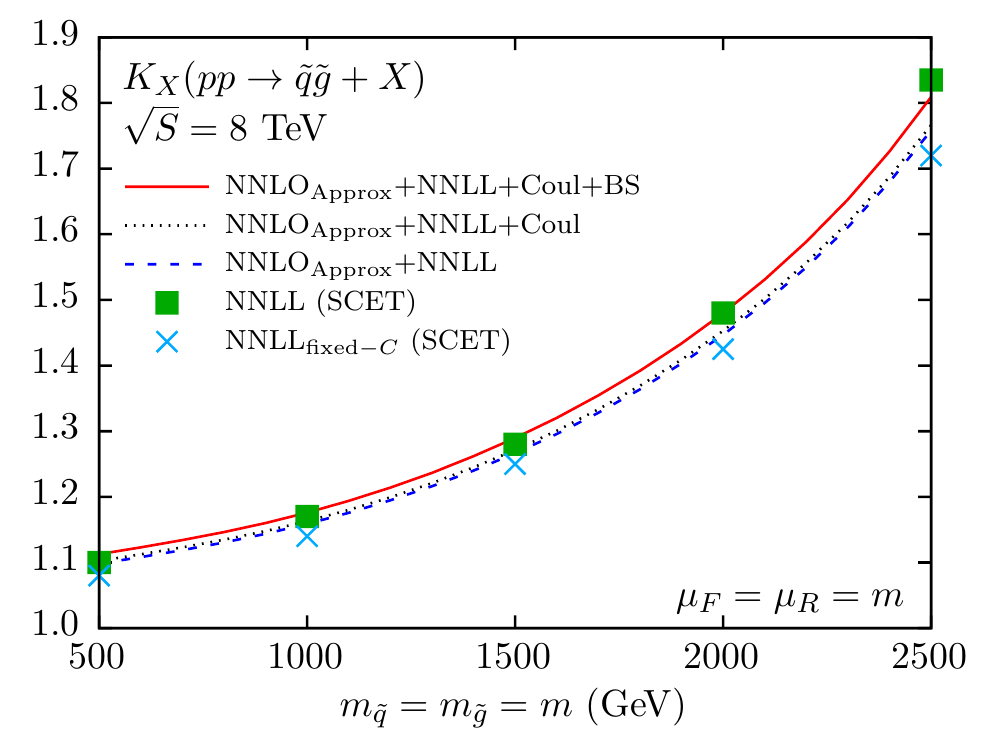}\includegraphics[width=0.47\textwidth]{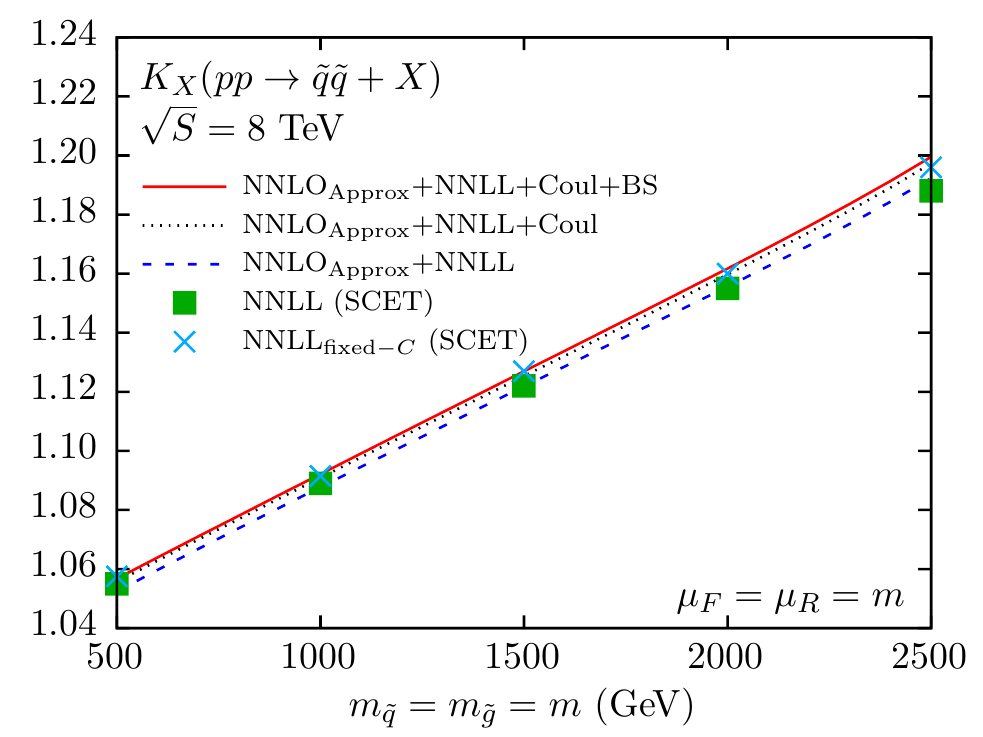}
	\caption{Comparison of the $K$-factors for squark and gluino production as obtained in the Mellin-space approach adopted in the present paper, compared with results obtained in soft-collinear effective theory (SCET). The cross sections correspond to a collider energy of $\sqrt{S} = 8$ TeV and the MSTW2008 NNLO PDF sets. The SCET results are taken from the corresponding plots in \cite{Beneke:2014wda}.}
	\label{fig:coulbskSCET}
\end{figure}

Notwithstanding these differences, the NNLO$_\mathrm{Approx}$+NNLL+ Coul+BS and NNLL (SCET) predictions agree very well, typically within a few percent, see \reffig{fig:coulbskSCET}. Larger differences of 10-20\% are observed only for very high sparticle masses in the $\glu\glu$ and $\tilde{q}\tilde{q}^*$ channels. Comparing 
``NNLO$_\mathrm{Approx}$+NNLL+Coul+BS'' and ``NNLL (SCET)'' with ``NNLO$_\mathrm{Approx}$+NNLL'' and  ``NNLL$_{\mathrm{fixed}-C}$ (SCET)'', respectively, 
we see that the contributions from Coulomb resummation and bound states are in general positive and increase the $K$-factors. An exception is observed in the $\squ\squ$-channel within the SCET formalism, where the inclusion of Coulomb resummation and bound state effects reduce the size of the higher-order corrections. 
The results, however, agree within the theoretical scale uncertainty.
 
\begin{figure}[h]
	\centering
	\includegraphics[width=0.47\textwidth]{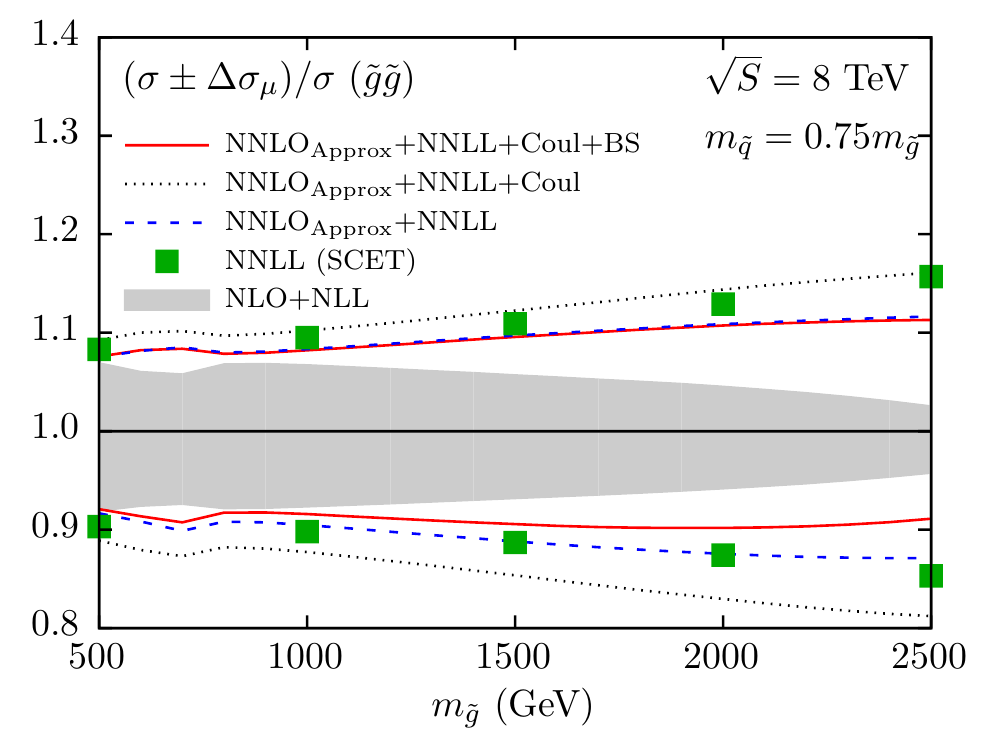}\includegraphics[width=0.47\textwidth]{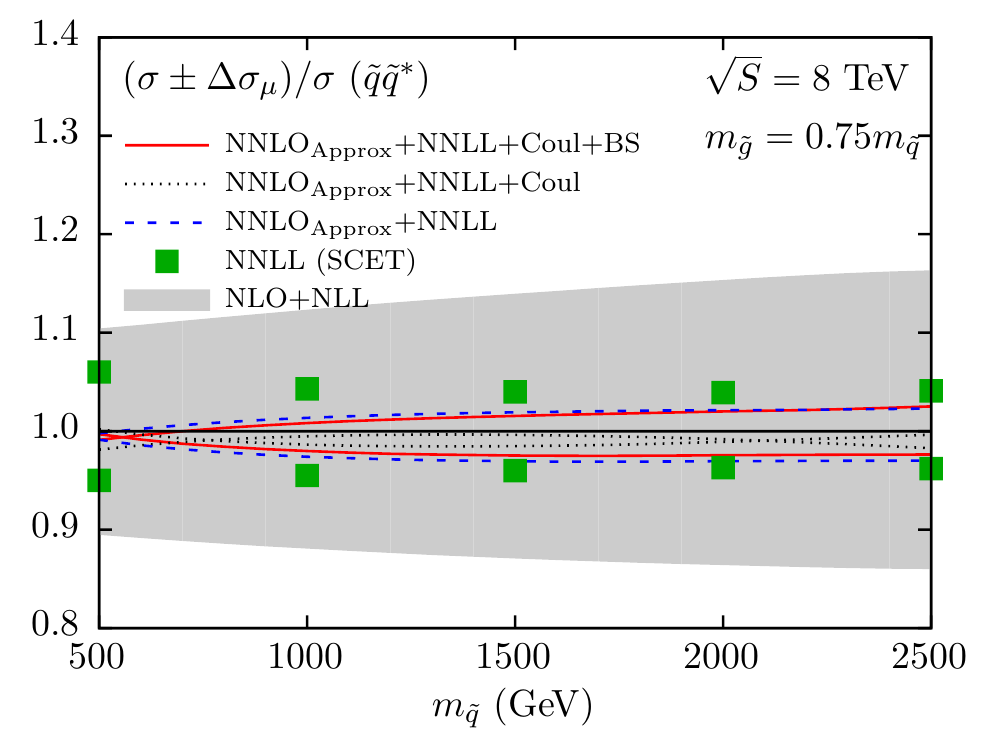}\\
	\includegraphics[width=0.47\textwidth]{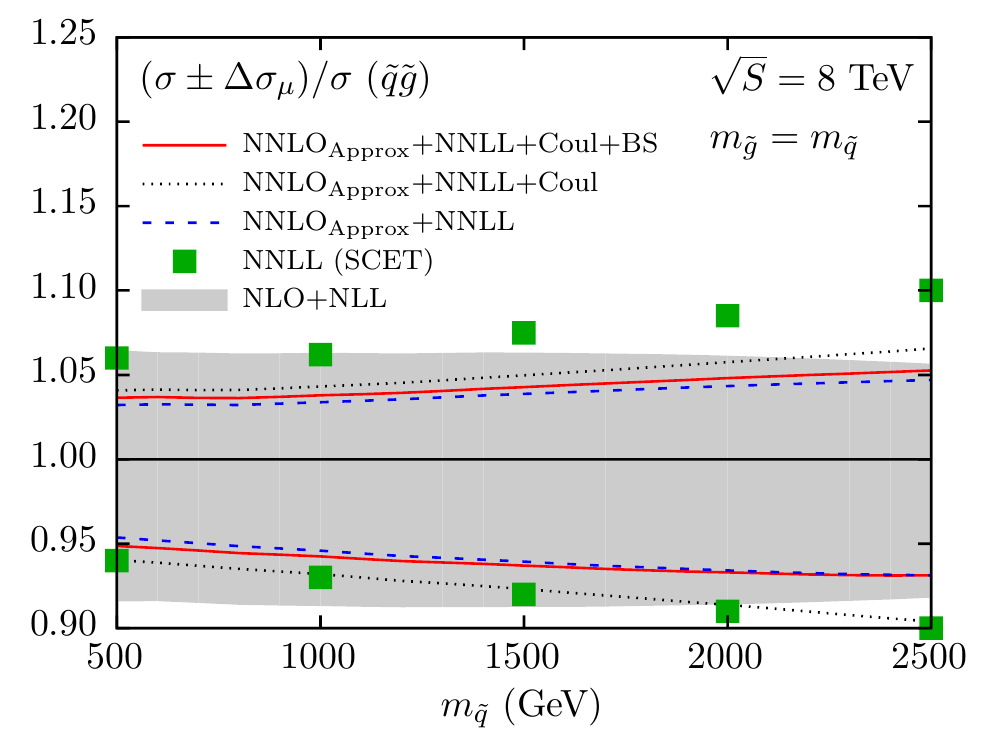}\includegraphics[width=0.47\textwidth]{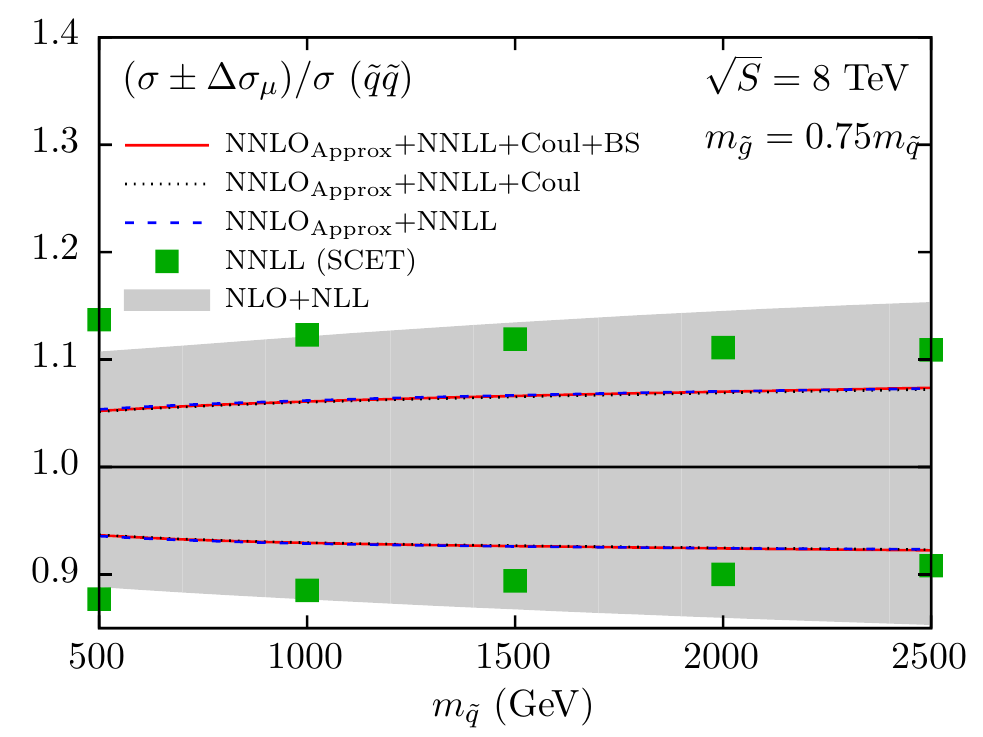}
	\caption{The relative scale uncertainty of the cross section prediction as obtained in the Mellin-space approach adopted in the present paper, compared with results obtained in soft-collinear effective theory (SCET). All scales have been varied by a factor of two about the central values. The cross sections correspond to a collider energy of $\sqrt{S} = 8$ TeV and the MSTW2008 NLO and NNLO PDF sets at NLL and NNLL accuracy, respectively. The SCET results are taken from the corresponding plots in \cite{Beneke:2014wda}.}
	\label{fig:coulbskmuSCET}
\end{figure}

The scale dependence of the cross sections, shown in \reffig{fig:coulbsmu} for our Mellin-space calculation, is not available for the SCET results. 
We can, however, quantify and compare the overall scale uncertainty of the cross section predictions, as obtained from varying all scales simultaneously by a factor of two about the central scale. This uncertainty is shown in  \reffig{fig:coulbskmuSCET} for the various predictions within the Mellin-space and SCET formalisms, respectively. Note that we have adopted unequal squark and gluino masses for the $\glu\glu$, $\squ\squ^*$, and $\squ\squ$ channels to enable a comparison with the SCET results of \cite{Beneke:2014wda}. 

Let us first discuss the scale uncertainty for the various Mellin-space predictions. In general we find that the NNLO$_\mathrm{Approx}$+NNLL scale uncertainty is smaller than at NLO+NLL. Exceptions are found for very large sparticle masses in the $\glu\glu$-channel as mentioned before and discussed in detail in \cite{Beenakker:2014sma}.  The inclusion of Coulomb resummation and bound-state contributions has a small effect on the overall scale uncertainty, which can be seen by comparing the NNLO$_\mathrm{Approx}$+NNLL and NNLO$_\mathrm{Approx}$+NNLL+Coul+BS predictions in  \reffig{fig:coulbskmuSCET}. Note that a slight increase in the scale dependence from the inclusion of Coulomb and bound state effects can be explained by the dependence on the Bohr and Coulomb scales, which are not present in our predictions at NLO+NLL and NNLO$_{\mathrm{Approx}}$+NNLL accuracy. Finally, comparing our Mellin-space results to SCET, we observe  a reasonable agreement in the estimate of the scale uncertainty, with our results falling within the NNLL (SCET) uncertainty band. This is not unexpected, as the additional variation of the soft scale, which is present in SCET, will in general lead to a larger scale uncertainty. 

\medskip

The results presented in figures~\ref{fig:coulbskSCET} and \ref{fig:coulbskmuSCET} show that the Mellin-space and SCET methods for threshold resummation lead to comparable results, in particular for the best predictions in the two approaches, NNLO$_\mathrm{Approx}$+NNLL+Coul+BS and NNLL (SCET), respectively, which include Coulomb resummation and bound-state effects. 

\section{Conclusions and outlook}\label{s:conclusion}
We have presented state-of-the-art SUSY-QCD predictions for squark and gluino production at the LHC, 
including NNLL threshold resummation, matched to approximate NNLO results, and taking into account Coulomb and bound-state corrections. 
The calculations have been done in the Mellin-space approach, and significantly improve on our previous NLO+NLL results. 

We find that the NNLL corrections are positive and sizeable, in particular for 
large squark and gluino masses of ${\cal O}({\rm TeV})$ and beyond.  The contributions from Coulomb resummation and bound 
states are also positive and result in a slight further increase of the cross-section predictions. The renormalization and factorization scale dependence  
decreases when including NNLL resummation, with a remaining overall scale uncertainty at NNLL accuracy of 5-10\%. The uncertainty of the squark and gluino  
cross-section predictions is now dominated by the parton distribution function uncertainty.  

We have compared our predictions with comparable calculations performed in the framework of soft-collinear effective theory. 
We find that the two methods lead to consistent results, in particular for the best predictions in the two approaches which include 
Coulomb resummation and bound-state effects. 

\medskip

The NNLL resummed cross section predictions for squarks and gluinos can be obtained from the computer code \texttt{NNLL-fast}. 
The code also provides the scale uncertainty obtained from the variation of the renormalisation, factorisation and Coulomb scales by a factor of two about the corresponding 
central scales, and the pdf and $\alpha_{\rm s}$ error. \texttt{NNLL-fast} can be downloaded at 
\begin{center}
 	\href{http://pauli.uni-muenster.de/~akule_01/nnllfast}{\texttt{http://pauli.uni-muenster.de/\textasciitilde{}akule\_01/nnllfast}}
\end{center}
and provides the theoretical basis to interpret current and future searches for supersymmetry at  the LHC.

\medskip

\noindent
{\it Note added:} While finalising this work we became aware of Ref.\,\cite{SCET}, which addresses the joint resummation of threshold and Coulomb corrections
for squark and gluino production at the LHC in the soft-collinear effective theory, and which provides a more comprehensive and detailed account of the work presented in Ref.\,\cite{Beneke:2013opa}.

\section*{Acknowledgments}
We thank Daniel Schwartl\"ander for performing multiple checks of results, and Christian Schwinn and the authors of Ref.~\cite{Beneke:2014wda} for valuable discussions. Part of this work has been performed on the High Performance Computing cluster PALMA maintained by the Center for Information Technology (ZIV) at WWU M\"unster, and on the high-performance computing resources funded by the Ministry of Science, Research and the Arts and the Universities of the State of Baden-W\"urttemberg, Germany, within the framework program bwHPC. We acknowledge support by the DFG through the Research Unit 2239 ``New Physics at the Large Hadron Collider''; by the BMBF under contract 05H15PMCCA; by the Institutional Strategy of the University of T\"ubingen (DFG, ZUK 63); by the Research Executive Agency (REA) of the European Union under the Grant Agreement number PITN-GA- 2012-316704 (HIGGSTOOLS); by the Foundation for Fundamental Research of Matter (FOM), programme 156, ``Higgs as Probe and Portal''; and by the Dutch National Organization for Scientific Research (NWO).

\bibliographystyle{JHEP}

\begin{thebibliography}{10}

\bibitem{Wess:1973kz}
	J.~Wess and B.~Zumino,
	{\it A Lagrangian Model Invariant Under Supergauge Transformations},
	Phys.\ Lett.\ {\bf B49} (1974) 52.

\bibitem{Wess:1974tw}
	J.~Wess and B.~Zumino,
	{\it Supergauge Transformations in Four-Dimensions},
	Nucl.\ Phys.\ {\bf B70} (1974) 39.

\bibitem{Fayet:1976et}
	P.~Fayet,
	{\it Supersymmetry and Weak, Electromagnetic and Strong Interactions},
	Phys.\ Lett.\ {\bf B64} (1976) 159.

\bibitem{Farrar:1978xj}
	G.~R.~Farrar and P.~Fayet,
	{\it Phenomenology of the Production, Decay, and Detection of New Hadronic States Associated with Supersymmetry},
	Phys.\ Lett.\ {\bf B76} (1978) 575.

\bibitem{Sohnius:1985qm}
	M.~F.~Sohnius,
	{\it Introducing Supersymmetry},
	Phys.\ Rept.\ {\bf 128} (1985) 39.

\bibitem{Martin:1997ns}
	S.~P.~Martin,
	{\it A Supersymmetry primer},
	Adv.\ Ser.\ Direct.\ High Energy Phys.\  {\bf 21} (2010) 1.

\bibitem{Aad:2015iea}
	G.~Aad {\it et al.} [ATLAS Collaboration],
	{\it Summary of the searches for squarks and gluinos using $ \sqrt{s}=8 $ TeV pp collisions with the ATLAS experiment at the LHC},
	JHEP {\bf 1510} (2015) 054.

\bibitem{Aad:2015baa}
	G.~Aad {\it et al.} [ATLAS Collaboration],
	{\it Summary of the ATLAS experiment's sensitivity to supersymmetry after LHC Run 1 - interpreted in the phenomenological MSSM},
        JHEP {\bf 1510} (2015) 134.

\bibitem{Aad:2015pfx}
  G.~Aad {\it et al.} [ATLAS Collaboration],
  {\it ATLAS Run 1 searches for direct pair production of third-generation squarks at the Large Hadron Collider},
  Eur.\ Phys.\ J.\ C {\bf 75} (2015) no.10,  510,
  Erratum: [Eur.\ Phys.\ J.\ C {\bf 76} (2016) no.3,  153].
 
\bibitem{Chatrchyan:2014goa}
  S.~Chatrchyan {\it et al.} [CMS Collaboration],
  {\it Search for supersymmetry with razor variables in pp collisions at $\sqrt{s}$=7 TeV},
  Phys.\ Rev.\ D {\bf 90} (2014) no.11,  112001.
  
\bibitem{Khachatryan:2015lwa}
	V.~Khachatryan {\it et al.} [CMS Collaboration],
	{\it Search for Physics Beyond the Standard Model in Events with Two Leptons, Jets, and Missing Transverse Momentum in pp Collisions at sqrt(s) = 8 TeV},
	JHEP {\bf 1504} (2015) 124.

\bibitem{Khachatryan:2016pup}
  V.~Khachatryan {\it et al.} [CMS Collaboration],
  {\it Search for direct pair production of scalar top quarks in the single- and dilepton channels in proton-proton collisions at $ \sqrt{s}=8 $ TeV},
  JHEP {\bf 1607} (2016) 027. 
 
\bibitem{Nilles:1983ge}
	H.~P.~Nilles,
	{\it Supersymmetry, Supergravity and Particle Physics},
	Phys.\ Rept.\  {\bf 110} (1984) 1.

\bibitem{Haber:1984rc}
	H.~E.~Haber and G.~L.~Kane,
	{\it The Search for Supersymmetry: Probing Physics Beyond the Standard Model},
	Phys.\ Rept.\  {\bf 117} (1985) 75.

\bibitem{Ellis:1983ed}
	J.~R.~Ellis and S.~Rudaz,
	{\it Search for Supersymmetry in Toponium Decays},
	Phys.\ Lett.\ {\bf B128} (1983) 248.

\bibitem{Beenakker:1994an}
	W.~Beenakker, R.~H\"opker, M.~Spira, and P.~M. Zerwas,
	{\it {Squark production at the Tevatron}},
	Phys.\ Rev.\ Lett.\ {\bf 74} (1995) 2905.

\bibitem{Beenakker:1995fp}
	W.~Beenakker, R.~H\"opker, M.~Spira, and P.~M. Zerwas,
	{\it {Gluino pair production at the Tevatron}},
	Z.\ Phys.\ {\bf C69} (1995) 163.

\bibitem{Beenakker:1996ch}
	W.~Beenakker, R.~H\"opker, M.~Spira, and P.~M. Zerwas,
	{\it {Squark and gluino production at hadron colliders}},
	Nucl.\ Phys.\ {\bf B492} (1997) 51.

\bibitem{Beenakker:1997ut}
	W.~Beenakker, M.~Kr\"amer, T.~Plehn, M.~Spira, and P.~M. Zerwas,
	{\it {Stop production at hadron colliders}},
	Nucl.\ Phys.\ {\bf B515} (1998) 3.

\bibitem{GoncalvesNetto:2012yt}
  D.~Goncalves-Netto, D.~Lopez-Val, K.~Mawatari, T.~Plehn and I.~Wigmore,
  {\it Automated Squark and Gluino Production to Next-to-Leading Order},
  Phys.\ Rev.\ D {\bf 87} (2013) no.1,  014002.

\bibitem{Hollik:2013xwa}
  W.~Hollik, J.~M.~Lindert and D.~Pagani,
  {\it On cascade decays of squarks at the LHC in NLO QCD,}
   Eur.\ Phys.\ J.\ C {\bf 73} (2013) 2410.

\bibitem{Gavin:2013kga}
  R.~Gavin, C.~Hangst, M.~Kr\"amer, M.~M\"uhlleitner, M.~Pellen, E.~Popenda and M.~Spira,
  {\it Matching Squark Pair Production at NLO with Parton Showers},
  JHEP {\bf 1310} (2013) 187.

\bibitem{Hollik:2012rc}
  W.~Hollik, J.~M.~Lindert and D.~Pagani,
  {\it NLO corrections to squark-squark production and decay at the LHC},
  JHEP {\bf 1303} (2013) 139.

\bibitem{Gavin:2014yga}
  R.~Gavin, C.~Hangst, M.~Kr\"amer, M.~M\"uhlleitner, M.~Pellen, E.~Popenda and M.~Spira,
  {\it Squark Production and Decay matched with Parton Showers at NLO},
  Eur.\ Phys.\ J.\ C {\bf 75} (2015) no.1,  29.

\bibitem{Degrande:2015vaa}
  C.~Degrande, B.~Fuks, V.~Hirschi, J.~Proudom and H.~S.~Shao,
  {\it Matching next-to-leading order predictions to parton showers in supersymmetric QCD},
  Phys.\ Lett.\ B {\bf 755} (2016) 82.

\bibitem{Hollik:2007wf}
	W.~Hollik, M.~Kollar and M.~K.~Trenkel,
	{\it Hadronic production of top-squark pairs with electroweak NLO contributions},
	JHEP {\bf 0802} (2008) 018.

\bibitem{Beccaria:2008mi}
	M.~Beccaria, G.~Macorini, L.~Panizzi, F.M.~Renard and C.~Verzegnassi,
	{\it Stop-antistop and sbottom-antisbottom production at LHC: A One-loop search for model parameters dependence},
	Int.\ J.\ Mod.\ Phys.\ {\bf A23} (2008) 4779.

\bibitem{Hollik:2008yi}
	W.~Hollik and E.~Mirabella,
	{\it Squark anti-squark pair production at the LHC: The Electroweak contribution},
	JHEP {\bf 0812} (2008) 087.

\bibitem{Hollik:2008vm}
	W.~Hollik, E.~Mirabella and M.~K.~Trenkel,
	{\it Electroweak contributions to squark-gluino production at the LHC},
	JHEP {\bf 0902} (2009) 002.

\bibitem{Mirabella:2009ap}
	E.~Mirabella,
	{\it NLO electroweak contributions to gluino pair production at hadron colliders},
	JHEP {\bf 0912} (2009) 012.

\bibitem{Hollik:2010vn}
	J.~Germer, W.~Hollik, E.~Mirabella and M.~K.~Trenkel,
	{\it Hadronic production of squark-squark pairs: The electroweak contributions},
	JHEP {\bf 1008} (2010) 023.

\bibitem{Germer:2014jpa}
  J.~Germer, W.~Hollik, J.~M.~Lindert and E.~Mirabella,
  {\it Top-squark pair production at the LHC: a complete analysis at next-to-leading order},
  JHEP {\bf 1409} (2014) 022.

\bibitem{Hollik:2015lha}
  W.~Hollik, J.~M.~Lindert, E.~Mirabella and D.~Pagani,
  {\it Electroweak corrections to squark-antisquark production at the LHC},
  JHEP {\bf 1508} (2015) 099.

\bibitem{Sterman:1986aj}
	G.~F. Sterman,
	{\it {Summation of Large Corrections to Short Distance Hadronic Cross-Sections}},
	Nucl.\ Phys.\ {\bf B281} (1987) 310.

\bibitem{Catani:1989ne}
	S.~Catani and L.~Trentadue,
	{\it {Resummation of the QCD Perturbative Series for Hard Processes}},
	Nucl.\ Phys.\ {\bf B327} (1989) 323.

\bibitem{Bonciani:1998vc}
	R.~Bonciani, S.~Catani, M.~L. Mangano, and P.~Nason, {\it {NLL Resummation of the Heavy-Quark Hadroproduction Cross-Section}},
	Nucl.\ Phys.\ {\bf B529} (1998) 424.

\bibitem{Contopanagos:1996nh}
	H.~Contopanagos, E.~Laenen, and G.~Sterman, {\it {Sudakov Factorization and Resummation}},
	Nucl.\ Phys.\ {\bf B484} (1997) 303.

\bibitem{Kidonakis:1998bk}
	N.~Kidonakis, G.~Oderda, and G.~Sterman, {\it {Threshold Resummation for Dijet Cross Sections}},
	Nucl.\ Phys.\ {\bf B525} (1998) 299.

\bibitem{Kidonakis:1998nf}
	N.~Kidonakis, G.~Oderda, and G.~Sterman, {\it {Evolution of Color Exchange in {QCD} Hard Scattering}},
	Nucl.\ Phys.\ {\bf B531} (1998) 365.

\bibitem{Kulesza:2008jb}
	A.~Kulesza and L.~Motyka,
	{\it Threshold resummation for squark-antisquark and gluino-pair production at the LHC},
	Phys.\ Rev.\ Lett.\  {\bf 102} (2009) 111802.

\bibitem{Kulesza:2009kq}
	A.~Kulesza and L.~Motyka,
	{\it Soft gluon resummation for the production of gluino-gluino and squark-antisquark pairs at the LHC},
	Phys.\ Rev.\ {\bf D80} (2009) 095004.

\bibitem{Beenakker:2009ha}
	W.~Beenakker, S.~Brensing, M.~Kr\"amer, A.~Kulesza, E.~Laenen and I.~Niessen,
	{\it Soft-gluon resummation for squark and gluino hadroproduction},
	JHEP {\bf 0912} (2009) 041.

\bibitem{Beenakker:2010nq}
	W.~Beenakker, S.~Brensing, M.~Kr\"amer, A.~Kulesza, E.~Laenen and I.~Niessen,
	{\it Supersymmetric top and bottom squark production at hadron colliders},
	JHEP {\bf 1008} (2010) 098.

\bibitem{Beenakker:2011fu}
	W.~Beenakker, S.~Brensing, M.~Kr\"amer, A.~Kulesza, E.~Laenen, L.~Motyka and I.~Niessen,
	{\it Squark and Gluino Hadroproduction},
	Int.\ J.\ Mod.\ Phys.\  {\bf A26} (2011) 2637.

\bibitem{Beneke:2010da}
	M.~Beneke, P.~Falgari and C.~Schwinn,
	{\it Threshold resummation for pair production of coloured heavy (s)particles at hadron colliders},
	Nucl.\ Phys.\ {\bf B842} (2011) 414.

\bibitem{Falgari:2012hx}
	P.~Falgari, C.~Schwinn and C.~Wever,
	{\it NLL soft and Coulomb resummation for squark and gluino production at the LHC},
	JHEP {\bf 1206} (2012) 052.

\bibitem{Beenakker:2011sf}
	W.~Beenakker, S.~Brensing, M.~Kr\"amer, A.~Kulesza, E.~Laenen and I.~Niessen,
	{\it NNLL resummation for squark-antisquark pair production at the LHC},
	JHEP {\bf 1201} (2012) 076.

\bibitem{Langenfeld:2012ti}
	U.~Langenfeld, S.~O.~Moch and T.~Pfoh,
	{\it QCD threshold corrections for gluino pair production at hadron colliders},
	JHEP {\bf 1211} (2012) 070.

\bibitem{Pfoh:2013edr}
	T.~Pfoh,
	{\it Phenomenology of QCD threshold resummation for gluino pair production at NNLL},
	JHEP {\bf 1305} (2013) 044
	[JHEP {\bf 1310} (2013) 090].

\bibitem{Beenakker:2013mva}
	W.~Beenakker {\it et al.},
	{\it Towards NNLL resummation: hard matching coefficients for squark and gluino hadroproduction},
	JHEP {\bf 1310} (2013) 120.

\bibitem{Beneke:2013opa}
	M.~Beneke, P.~Falgari, J.~Piclum, C.~Schwinn and C.~Wever,
	{\it Higher-order soft and Coulomb corrections to squark and gluino production at the LHC},
	PoS RADCOR {\bf 2013} (2013) 051.

\bibitem{Beneke:2014wda}
	M.~Beneke, P.~Falgari, J.~Piclum, C.~Schwinn and C.~Wever,
	{\it Higher-order soft and Coulomb corrections to squark and gluino production at the LHC},
	PoS LL {\bf 2014} (2014) 060.

\bibitem{Beenakker:2014sma}
	W.~Beenakker, C.~Borschensky, M.~Kr\"amer, A.~Kulesza, E.~Laenen, V.~Theeuwes and S.~Thewes,
	{\it NNLL resummation for squark and gluino production at the LHC},
	JHEP {\bf 1412} (2014) 023.

\bibitem{Broggio:2013cia}
	A.~Broggio, A.~Ferroglia, M.~Neubert, L.~Vernazza and L.~L.~Yang,
	{\it NNLL Momentum-Space Resummation for Stop-Pair Production at the LHC},
	JHEP {\bf 1403} (2014) 066.

\bibitem{Beenakker:2016gmf}
	W.~Beenakker, C.~Borschensky, R.~Heger, M.~Kr\"amer, A.~Kulesza and E.~Laenen,
	{\it NNLL resummation for stop pair-production at the LHC},
	JHEP {\bf 1605} (2016) 153.

\bibitem{Fadin:1990wx}
	V.~S.~Fadin, V.~A.~Khoze and T.~Sj\"ostrand,
	{\it On the Threshold Behavior of Heavy Top Production},
	Z.\ Phys.\ {\bf C48} (1990) 613.

\bibitem{Kwong:1990iy}
	W.~k.~Kwong,
	{\it Threshold production of $t \bar{t}$ pairs by $e^{+} e^{-}$ collisions},
	Phys.\ Rev.\ {\bf D43} (1991) 1488.

\bibitem{Strassler:1990nw}
	M.~J.~Strassler and M.~E.~Peskin,
	{\it The Heavy top quark threshold: QCD and the Higgs},
	Phys.\ Rev.\ {\bf D43} (1991) 1500.

\bibitem{Fadin:1987wz}
  V.~S.~Fadin and V.~A.~Khoze,
  {\it Threshold Behavior of Heavy Top Production in e+ e- Collisions}, 
  JETP Lett.\  {\bf 46} (1987) 525
   [Pisma Zh.\ Eksp.\ Teor.\ Fiz.\  {\bf 46} (1987) 417].

\bibitem{Hoang:2000yr}
  A.~H.~Hoang {\it et al.},
  {\it Top - anti-top pair production close to threshold: Synopsis of recent NNLO results},
  Eur.\ Phys.\ J.\ direct C {\bf 3} (2000) 1.
    
    
\bibitem{Beneke:1999qg}
  M.~Beneke, A.~Signer and V.~A.~Smirnov,
  {\it Top quark production near threshold and the top quark mass},
  Phys.\ Lett.\ B {\bf 454} (1999) 137.

\bibitem{nnll-fast}    
	\href{http://pauli.uni-muenster.de/~akule_01/nnllfast}{\texttt{http://pauli.uni-muenster.de/\textasciitilde{}akule\_01/nnllfast}}
  
\bibitem{Botts:1989kf}
        J.~Botts and G.~Sterman,
        {\it Hard Elastic Scattering In QCD: Leading Behavior},
        Nucl.\ Phys.\ {\bf B325} (1989) 62.

\bibitem{Kidonakis:1997gm}
	N.~Kidonakis and G.~F.~Sterman,
	{\it Resummation for QCD hard scattering},
	Nucl.\ Phys.\ {\bf B505} (1997) 321.

\bibitem{Beneke:2009rj}
	M.~Beneke, P.~Falgari and C.~Schwinn,
	{\it Soft radiation in heavy-particle pair production: All-order colour structure and two-loop anomalous dimension},
	Nucl.\ Phys.\ {\bf B828} (2010) 69.

\bibitem{Falgari:2012sq}
  P.~Falgari, C.~Schwinn and C.~Wever,
  {\it Finite-width effects on threshold corrections to squark and gluino production},
  JHEP {\bf 1301} (2013) 085.

\bibitem{Pineda:2006ri}
	A.~Pineda and A.~Signer,
	{\it Heavy Quark Pair Production near Threshold with Potential Non-Relativistic QCD},
	Nucl.\ Phys.\ {\bf B762} (2007) 67.

\bibitem{Kiyo:2008bv}
	Y.~Kiyo, J.~H.~K\"uhn, S.~Moch, M.~Steinhauser and P.~Uwer,
	{\it Top-quark pair production near threshold at LHC},
	Eur.\ Phys.\ J.\ {\bf C60} (2009) 375.

\bibitem{Kauth:2011bz}
        M.~R. Kauth, A.~Kress and J.~H. K\"uhn, 
        {\it Gluino-Squark Production at the LHC: The Threshold},
        JHEP {\bf 1112} (2011) 104.

\bibitem{Kauth:2011vg}
        M.~R. Kauth, J.~H. K\"uhn, P.~Marquard and M.~Steinhauser, 
        {\it Gluino Pair Production at the LHC: The Threshold},
        Nucl.\ Phys.\ {\bf B857} (2012) 28.

\bibitem{Hagiwara:2009hq}
        K.~Hagiwara and H.~Yokoya,
        {\it Bound-state effects on gluino-pair production at hadron colliders},
        JHEP {\bf 0910} (2009) 049.

\bibitem{Beneke:2009ye}
	M.~Beneke, M.~Czakon, P.~Falgari, A.~Mitov and C.~Schwinn,
	{\it {Threshold expansion of the $gg(q\bar{q}) \to Q\bar{Q} + X$ cross section at ${\cal O}(\alpha_s^4)$}},
	Phys.\ Lett. {\bf B690} (2010) 483.

\bibitem{Baernreuther:2013caa}
  P.~B\"arnreuther, M.~Czakon and P.~Fiedler,
  {\it Virtual amplitudes and threshold behaviour of hadronic top-quark pair-production cross sections,}
  JHEP {\bf 1402} (2014) 078.

\bibitem{PSRtalk}
	C.\ Schwinn, talk at the Parton Shower and Resummation Workshop, 04-06 July 2016, Paris, France.\\
	https://indico.cern.ch/event/450053/
	
\bibitem{Beneke:2005hg}
	M.~Beneke, Y.~Kiyo and K.~Schuller,
	{\it Third-order Coulomb corrections to the S-wave Green function, energy levels and wave functions at the origin},
	Nucl.\ Phys.\ {\bf B714} (2005) 67.

\bibitem{Catani:1996yz}
	S.~Catani, M.~L.~Mangano, P.~Nason and L.~Trentadue,
	{\it The Resummation of soft gluons in hadronic collisions},
	Nucl.\ Phys.\ {\bf B478} (1996) 273.

\bibitem{Agashe:2014kda}
  K.~A.~Olive {\it et al.} [Particle Data Group Collaboration],
  {\it Review of Particle Physics},
  Chin.\ Phys.\ C {\bf 38} (2014) 090001.

\bibitem{Butterworth:2015oua}
  J.~Butterworth {\it et al.},
  {\it PDF4LHC recommendations for LHC Run II},
  J.\ Phys.\ G {\bf 43} (2016) 023001.

\bibitem{Kulesza:2002rh}
	A.~Kulesza, G.~F.~Sterman and W.~Vogelsang,
	{\it Joint resummation in electroweak boson production},
	Phys.\ Rev.\ {\bf D66} (2002) 014011.

\bibitem{Beenakker:2015rna}
  W.~Beenakker, C.~Borschensky, M.~Kr\"amer, A.~Kulesza, E.~Laenen, S.~Marzani and J.~Rojo,
  {\it NLO+NLL squark and gluino production cross-sections with threshold-improved parton distributions},
  Eur.\ Phys.\ J.\ C {\bf 76} (2016) no.2,  53.

\bibitem{Hagiwara:2008df}
	K.~Hagiwara, Y.~Sumino and H.~Yokoya,
	{\it Bound-state Effects on Top Quark Production at Hadron Colliders},
	Phys.\ Lett.\ {\bf B666} (2008) 71.

\bibitem{prospino}
	W.~Beenakker, R.~H\"opker, and M.~Spira, {\it\texttt{PROSPINO}: A Program for the production of supersymmetric particles in next-to-leading order QCD},
	[\href{http://arxiv.org/abs/hep-ph/9611232}{\tt hep-ph/9611232}].
	See
	\url{http://www.thphys.uni-heidelberg.de/~plehn/index.php?show=prospino} or
	\url{http://tiger.web.psi.ch/prospino/}, 1996.

\bibitem{Beneke:2011mq}
	M.~Beneke, P.~Falgari, S.~Klein and C.~Schwinn,
	{\it Hadronic top-quark pair production with NNLL threshold resummation},
	Nucl.\ Phys.\ {\bf B855} (2012) 695.

\bibitem{SCET}
M.~Beneke, J.~Piclum, C.~Schwinn, C.~Wever,  
{\it NNLL soft and Coulomb resummation for squark and gluino  production at the LHC}, 
to be published.  

\end{thebibliography}
\providecommand{\href}[2]{#2}\begingroup\raggedright\endgroup

\end{document}